\documentclass[modern,twocolumn,pdftex]{aastex631}        \newif\ifdraft \drafttrue
\ifdraft \else \newcommand{\submitjournal}[1]{\relax} 
               \newcommand{\correspondingauthor}[1]{\relax} 
\fi
\submitjournal{AJ}
\usepackage{graphicx}
\usepackage{xcolor}
\usepackage{upgreek}
\usepackage{amsmath}	
\ifdraft
   \usepackage{hyperref}
   \hypersetup{
               colorlinks=true,
               linkcolor=black,
               filecolor=black,
               citecolor=black,
               urlcolor=blue
              }
     \newcommand{\web}[1]{\Blb{\url{#1}}}
\else
     \newcommand{\web}[1]{#1}
\fi

\newcommand{\PIMA}{$\cal P\hspace{-0.067em}I\hspace{-0.067em}M\hspace{-0.067em}A$ }

\newcommand{\nnntab}[2]{ \multicolumn{3}{#1}{#2} }

\definecolor{Dred}{rgb}{0.312,0.070,0.070}
\definecolor{Dblue}{rgb}{0.070,0.070,0.312}
\definecolor{Dgreen}{rgb}{0.070,0.312,0.070}
\definecolor{Db}{rgb}    {0.050,0.0,0.320}

\newcommand{\Blb}[1]{\textcolor{Dblue}{\bf #1}}

\newcommand{\beq}{ \begin{eqnarray} }
\newcommand{\eeq}[1]{\label{#1}\end{eqnarray}}
\newcommand{\eeqn}{ \nonumber \end{eqnarray} }
\newcommand{\Frac}[2]{\frac{\displaystyle\strut #1}{\displaystyle\strut #2} }
\newcommand{\vex}{\vspace{1ex}}

\renewcommand{\tau}{\uptau}

\newcommand{\Corr}{\rm Corr}
\newcommand{\Var}{\rm Var}

\newcounter{note}
\let\oldmarginpar\marginpar
\renewcommand\marginpar[1]{\-\oldmarginpar[\raggedleft\footnotesize #1]%
{\raggedright\footnotesize #1}}

\newcommand{\Number}[1]{\ifnum#1<10\relax0\number#1\else\number#1\fi}
\newcommand{\isodate}{
\count151=\time
\divide\count151 by 60
\count151=\count151
\multiply\count151 by 60
\count152=\time
\advance\count152 by -\count151
\divide\count151 by 60
\count152=\count151
\multiply\count151 by 60
\count153=\time
\advance\count153 by -\count151
\Number{\year}.\Number{\month}.\Number{\day}--\Number{\count152}:\Number{\count153}
}

\received{Apil 12, 2024}
\revised{}
\accepted{}

\begin{document}

\title{Radioastrometry at different frequencies}
\ifdraft \relax \else \correspondingauthor{Leonid Petrov} \fi
\email{Leonid.Petrov@nasa.gov}

\author[0000-0001-9737-9667]{Leonid Petrov}
\affil{NASA Goddard Space Flight Center \\
Code 61A1, 8800 Greenbelt Rd, Greenbelt, 20771 MD, USA}


\begin{abstract}

   Very long baseline interferometry (VLBI) technique allows us to 
determine positions of thousands of radio sources using the absolute
astrometry approach. I have investigated the impact of a selection of 
observing frequencies in a range from 2 to 43~GHz in single-band, 
dual-band, and quad-band observing modes on astrometric results. 
I processed seven datasets in a range of 72 thousands to 6.9 million 
observations, estimated source positions, and compared them. I found 
that source positions derived from dual-band, quad-band, and 23.6~GHz 
single-band data agree at a level below 0.2~mas. Comparison of independent 
datasets allowed me to assess the error level of individual catalogues: 
0.05--0.07~mas per position component. Further comparison showed that 
individual catalogues have systematic errors at the same level. Positions 
from 23.6~GHz single-band data show systematic errors related to the 
residual ionosphere contribution. Analysis of source positions differences 
revealed systematic errors along jet directions at a level of 0.09~mas. 
Network related systematic errors affect all the data regardless of frequency. 
Comparison of position estimates allowed me to derive the stochastic error 
model that closes the error budget. Based on collected evidence, I made 
a conclusion that a development of frequency-dependent reference frames 
of the entire sky is not warranted. In most cases dual-band, quad-band, 
and single-band data at frequency 22~GHz and higher can be used 
interchangeably, which allows us to exploit the strength of a specific 
frequency setup for given objects. Mixing observations at different 
frequencies causes errors not exceeding 0.07~mas.

\end{abstract}


\section{Introduction}

   The method of very long baseline interferometry (VLBI) first proposed
by \citet{r:mat65} turned out a powerful tool for geodesy and astronomy. 
Analysis of VLBI regular campaigns that run since 1980s allowed us to 
determine positions and velocities of observing stations, time series of 
the Earth orientation parameters, and coordinates of observed sources,
including artificial satellites. VLBI hardware allows us to operate at 
frequencies from 20~MHz to 230~GHz, and most of the stations have the 
capability to tune observing frequencies or observe several frequencies 
simultaneously within a certain range. Therefore, we have a freedom to 
select a frequency range when plan observations. A question emerges:
which frequency range should we select? I will limit further discussion 
to a case of absolute astrometry based on processing total group 
delays. Since path delay is proportional to the dot product of 
the baseline vector and the unit source vector, source position accuracy 
does not depend on frequency directly. In the absence of systematic errors,
it is proportional to the precision of derived group delays and reciprocal 
to the baseline length. In that ideal case if one wants to reach the 
highest positional accuracy, observations need to be done at a network with 
longest baselines, at frequencies where a given source has the strongest 
flux density, and using such a setup that provides the widest spanned 
bandwidth.

  However, systematic errors often dominate the error budget. The most 
relevant systematic errors are those that are caused by mismodeling path 
delay in the neutral atmosphere, mismodeling the ionospheric contribution, 
errors caused by source structure, and a core shift.

  Path delay in the neutral atmosphere virtually does not depend on 
frequency in a range of 1 to 1000~GHz. But even if path delay itself does 
not depend on frequency, our ability to estimate residual path delay with 
respect to the a~priori does depend. A robust estimation of the atmospheric 
path delay in zenith direction requires observing sources at both high and 
low elevations. \citet{r:mcm94} found that the systematic errors reach the 
minimum when observations at elevation angles $7^\circ$ are included. 
At frequencies above 15~GHz atmospheric opacity becomes noticeable. If 
the atmospheric opacity is 0.5 in the zenith direction, it can reach 
2.5--3.0 at $10^\circ$ elevation. That means flux density in the zenith 
direction is attenuated by a factor of 1.4, while at $10^\circ$ elevation
it is attenuated by a factor of 15--20. We may not be able to detect 
a source because of such a strong attenuation.

  The impact of the ionosphere is reciprocal to the square of frequency.
One can neglect the ionospheric contribution at frequencies 43~GHz and
above. At frequencies 10~GHz and lower, the ionospheric contribution,
if not properly modeled, dominates the error budget. The solution of
the problem was known for decades: to observe at two widely separated
frequencies and form an ionosphere-free linear combination of observables.
This approach works remarkably well at frequencies 2--10~GHz: the residual
ionospheric contribution due to higher-order terms in the expansion of the 
dispersiveness on frequency is not detectable \citep{r:iono2nd}. Later,
the systems that simultaneously record four frequencies were 
developed \citep{r:vgos}. 

  Over 90\% observed sources exhibit structure detectable at images.
The contribution of source structure to delay causes systematic errors.
This effect was first studied in detail by \citet{r:tho80}. When we know 
the source brightness distribution, we can compute the contribution 
to delay and correct the data. The first attempt to apply this contribution 
was done by \citet{r:zep88}. Although it was demonstrated by \citet{r:cha02} 
that a massive application of structure maps to correct for structure effects
is feasible, and it improves results of data analysis, so far, this approach
did not go beyond demonstrations, and up to now, the source 
structure contribution is not modeled in a routine VLBI data analysis. 
\citet{r:pla16} have shown through simulations that the contribution 
of source structure to delay at 8~GHz is in a range of 10--80 $\mu$as 
for {\it most} of the sources. Analysis of 
VLBI observations of 29~active galactic nuclei (AGNs) observed under 
MOJAVE program \citep{r:mojave2} at 15~GHz reported in \citet{r:gaia3} 
showed results that are consistent with simulations: applying the source 
structure contribution from images changed positions in a range from 
0.01 to 2.40 mas with the median 0.06~mas. \citet{r:moor11} presented
evidence of weak correlation between jet directions and directions
of observed AGN proper motion. If to scale a source image reciprocal
to frequency, the contribution of source structure to delay will also 
be scaled reciprocal to frequency, i.e. reduced. However, one should 
keep in mind that it is the asymmetry of the core region at scales
comparable with the resolution that affects the source structure 
contribution the most. This asymmetry does not vanish even at frequencies 
as high  as 230~GHz \citep{r:3c279_eht} and may be more profound at 
high frequencies. It should also be noted that source structure changes 
are more significant at high frequencies. 

  Due to synchrotron self-absorption, the AGN core center is observed 
at a location where the optical depth is close to one. That location 
depends on frequency $f$. In a case when the energy of the magnetic 
field and relativistic particles is approximately equal 
(the equipartition condition), this dependence is $f^{-1}$ 
\citep{r:lob98}. Observations confirm that it is a common situation 
\citep{r:kov08,r:sok11,r:abe18}, however examples of deviations of the 
power law from -1 are also known. When the core-shift depends on 
frequency as $f^{-1}$, the contribution to fringe phase becomes 
frequency independent, and since group delay is a partial derivative of 
phase over frequency, it does not affect group delay \citep{r:por09}. 
At the same time, it affects fringe phase and phase delay. A detailed 
study of a sample of thousands AGNs, revealed that the core-shift is 
variable \citep{r:pla19}, and the deviation from the equi-partition 
is associated with a flaring activity, typically on a time scale 
of several years. Therefore, in a case if a source is in the 
equipartition state, the core-shift does not affect absolute source 
positions derived from analysis of group delays, although it does affect 
positions derived with a method of differential astrometry based on 
analysis of differential phases. When the equipartition condition is 
violated, for instance within several years of a flare, the variable 
core-shift affects group delays as well, and this contribution is 
reciprocal to frequency.

  Scintillations in the interstellar medium cause broadening of source
images. As a result, the correlated flux density at long baselines is 
reduced. Since position accuracy is reciprocal to the projected baseline 
length, a drop of correlated flux density, especially if it falls below 
the detection limit, affects detrimentally the position accuracy. 
\citet{r:scat1,r:scat2} found that scattering is significant in the 
vicinity of the Galactic plane and in areas with a high density of 
the interstellar medium. This effect is reciprocal to the square of 
frequency and it barely affects data at 22~GHz, but it may severally 
affect observations at 2--8~GHz.

\subsection{Problem statement and approach}

  As we see, the impact of four effects, atmospheric path delay, contribution 
of the ionosphere, source structure, and core-shift depends on the choice of
observed frequency or frequencies. What are trade-offs? One can consider
two extreme cases: 1)~a frequency-dependent bias in position estimates is 
unique to each band and has to be measured; 2)~the frequency-dependent bias
is negligible. Realistically, we can assume the truth is between these
extremes, but is it close to case 1 or to case 2? An answer to this question
has a profound impact on strategy of radioastrometry. If the 
frequency-dependent bias is significant, then a celestial reference frame 
to each frequency band or each combination of frequency bands needs be 
constructed. This would require significant resources that have to be taken 
by displacing other projects. From the other hand, if the frequency-dependent 
bias is negligible, then the radio celestial reference frame is frequency 
independent as well, and we can combine observations at different frequencies 
to improve the position accuracy.

  Let us summarize five sources of frequency-dependent systematic errors:

\begin{itemize}
   \item Impact of path delay in the neutral atmosphere: disfavors
         frequencies above 20~GHz since at higher frequencies observations 
         at low elevations are downweighted or lost due to high atmospheric 
         opacity;

   \item impact of path delay in the ionosphere: disfavors single-band 
         observations below 43~GHz;

   \item impact of the source structure: favors higher frequencies with 
         reservations;

   \item impact of the core-shirt: favors higher frequencies since the 
         core-shift is reduced with frequency.

   \item impact of the interstellar medium: favors high frequencies;
\end{itemize}

  It should be also noted that at higher frequencies in general, sources 
are weaker, sensitivity of radiotelescopes is lower, and therefore, 
precision of group delays is worse.

  Had we had a precise and reliable model of the error budget of all these 
contributions, we could solve this problem analytically. Although there is 
a certain progress in this direction, we are still far from claiming that 
we do quantitatively understand the error budget theoretically. Therefore, 
I consider another approach: processing existing multi-frequency VLBI data 
suitable for absolute astrometry analysis and assessing  differences 
in source position estimates. Analysis of the magnitude of 
frequency-dependent position biases will help us to answer the questions 
formulated at the beginning of this subsection: what are the trade-offs 
in the frequency selection for absolute astrometry programs, when should 
we process observations at different frequencies combined, when should 
we process them separately, and finally, whether efforts for constructing 
celestial reference frames for each individual frequency or a frequency 
combination are justified by evidence.

\section{Observations and data analysis}

  I have processed seven VLBI datasets. They include 1)~a 48~h
campaign of observations with frequency switching between 2.2/8.4 GHz and 
4.1/7.4 GHz with Very Long Baseline Array (VLBA); 2)~a campaign of 281 
experiments at 4.1/7.4~GHz with VLBA; 3)~a campaign of 270 experiments 
at 2.2/8.4 GHz with VLBA; 4)~a campaign of 2259 experiments at 2.2/8.4~GHz 
run by the International VLBI Service for geometry and astrometry (IVS); 
5)~a campaign of 153 experiments at 3.0/5.2/6.4/10.2~GHz at the network 
of IVS stations equipped with quad-band receivers; 6)~a campaign of 
90 experiments with VLBA at 23~GHz; and 7)~a campaign of 8 experiments 
with VLBA at 43~GHz.

  The original records of voltage from radio telescope receivers were 
correlated, and time series of cross- and auto- correlation data
have been computed forming so-called Level 1 data \citep{r:difx1,r:difx2}. 
Then group delays were evaluated from cross- and auto- correlation data using 
the fringe-fitting procedure, with either \PIMA \citep{r:vgaps} or Fourtit 
software\footnote{https://www.haystack.mit.edu/haystack-observatory-postprocessing-system-hops/}.
I ran the fringe fitting data analysis using \PIMA for all the data, except
IVS campaigns. For the latter two campaigns I used group delays
derived by Fourfit and stored in geodetic database files that are 
available at the NASA CDDIS data archive\footnote{https://cddis.nasa.gov}.

  Further astrometric data analysis was performed using group delays. 
That involved several steps: computation of theoretical path delays and 
forming small differences between observed and modeled delays; preprocessing 
that includes outliers elimination, weight update, and identifying clock 
breaks; and parameter estimation with least squares using all the 
data of each observing campaign.

\subsection{Data reduction and parameter estimation}

  I processed the data with the state-of-the art theoretical model used in
prior works, for instance in \citet{r:vgaps,r:wfcs}. In general, it follows
the so-called IERS Conventions \citep{r:iers2010}, with a number of 
improvements. Of them, the following are relevant for the present study:
Galactic aberration was accounted for, a~priori slant path delays were
computed by a direct integration of equations of wave propagation through
the heterogeneous atmosphere \citep{r:padel} using the output of NASA 
numerical weather model GEOS-FPIT \citep{r:geos18}, and the ionospheric 
contribution computed from the GNSS global ionospheric model 
CODE \citep{r:schaer99} for processing single-band delays with three
important modifications: the nominal height of the ionosphere was increased
by 56.7~km, elevation for the ionospheric mapping function was scaled 
by 0.9782, and the total electron contents was scaled by 0.85.
A thorough discussion of the impact of these modifications in given in
\citet{r:sba}.

  Group delays or ionosphere-linear combinations of observables at two or
more bands were fitted to the parametric model using least squares.
The parameters were partitioned into three classes: global parameters that
were estimated using the entire dataset, local parameters that were estimated
for each observing session, and segmented parameters that were estimated for 
each station for an interval of time that is shorter than an observing 
session. The parametric model included estimation of the following global 
parameters: right ascensions and declinations of all the sources, station 
positions at the reference epoch, station velocities, sine and cosine 
components of harmonic site position variations \citep{r:harpos}, and the 
non-linear motion of some stations with breaks due to seismic activity 
modeled with B-splines with multiple knots at epochs of seismic events. 
Polar motion, UT1, their rate of change, as well as nutation daily offsets 
were estimated as local parameters. Clock function, atmospheric path delay, 
and the tilt of the symmetry axis of the refractivity field for all the 
stations were modeled as an expansion over the B-spline basis of the 
1st degree. These coefficients were estimated as segmented parameters. 
The span between knots was 60~minutes for clock function, 20~minutes for 
the atmospheric path delay in zenith direction, and 6~hours for tilt angles. 
No-net-rotation constraints were imposed on a subset of source position 
estimates in order to find a solution of a linear problem of the incomplete 
rank (see \citet{r:wfcs} for more details). The subset of sources used for 
constraints included all the sources from the ICRF1 catalogue \citep{r:icrf1} 
that had at least 200~usable observations in processed campaigns.
 
\subsection{Error analysis}

  The source position uncertainties were derived from the uncertainties of 
group delays following the law of error propagation. An uncertainty of group 
delay is computed based in the signal to noise ratio of fringe amplitude. 
The noise of fringe visibilities is determined by Fourfit and 
\PIMA differently. Fourfit computes the noise theoretically from the number
of recorded bits. \PIMA uses a more sophisticated algorithm: it computes the 
noise level from the visibility data themselves. At the final stages of the 
fringe fitting procedure, the least square adjustment, \PIMA applies the 
additive reweighting procedure: it finds additive phase weight corrections 
for each observation that, being added in quadrature, makes the ratio of 
the weighted sum of residual phases to their mathematical expectations 
close to unity. 

  A similar procedure was performed for an update of group delay 
uncertainties. The extra variance was computed for each baseline and each 
experiment during a preprocessing stage of data analysis. This variance 
was added in quadrature to the reported group delay uncertainties, and
these inflated group delay uncertainties were used as reciprocal weights.
The ratios of the weighted sum of group delay residuals to their mathematical 
expectations were made to be close to unity for each baseline and each 
experiment. The algorithm for additive variance computation can be found 
in \citet{r:wfcs}.

  To assess the validity of reported uncertainties, I ran two additional 
decimation solutions for each dataset. Observations of each source were 
sorted in the chronological order and split into segments of 32 observations.
Observations were marked as belonging to odd or even segments 
{\tt oooooo eeeeee oooooo} \ldots, where letters {\tt o} and {\tt e} denote
odd and even segments respectively. Odd segments were downweighted by a factor 
of 1000 in the first decimation solution, and even segments were downweighted 
in the second solution. Then the estimates of source positions from odd and 
even decimation solutions have been compared. I computed arc lengths between 
position estimates and uncertainties of these arcs assuming the source 
position estimates are independent. First, I converted uncertainties of 
source positions over right ascension $\sigma_\alpha$, over declination
$\sigma_\delta$, and correlations between these uncertainties to the 
semi-major $\sigma_{\rm maj}$ and semi-minor axes $\sigma_{\rm min}$ of 
the error ellipse and the position angles $\theta$ of the semi-major axes 
counted from the North celestial pole counter-clockwise:

\begin{widetext}
\beq
   \begin{array}{lcl}
   \theta & = & \Frac{1}{2} \arctan{ \Frac{2 \, \Corr \, \sigma_\alpha\cos\delta \, \sigma_\delta}
                               {\sigma^2_\alpha \cos{\delta}^2 - \sigma^2_\delta} } 
   \vex \\
   \sigma_{\rm maj}^2 & = & \sigma^2_\alpha \cos{\delta}^2 + \sigma^2_\delta + 
                            \sqrt{ (\sigma^2_\alpha \cos{\delta}^2 - \sigma^2_\delta)^2 +
                                   4 \, (\Corr \, \sigma_\alpha\cos{\delta} \, \sigma_\delta)^2}/2
   \vex \\
   \sigma_{\rm min}^2 & = & \sigma^2_\alpha \cos{\delta}^2 + \sigma^2_\delta - 
                            \sqrt{ (\sigma^2_\alpha \cos{\delta}^2 - \sigma^2_\delta)^2 +
                                   4 \, (\Corr \, \sigma_\alpha\cos{\delta} \, \sigma_\delta)^2}/2
   \end{array}
\eeq{e:e1}
 
  Then the uncertainty of an arc between position 1 and position 2 is
\beq
     \sigma^2_a = \Frac{1 + \tan^2(\theta_1 - \phi)}
                       {1 + \frac{\sigma^2_{1,\rm maj}}{\sigma^2_{1,\rm min}} \tan^2(\theta_1 - \phi)} \, \sigma^2_{1,\rm maj} \, 
                  \: + \:
                  \Frac{1 + \tan^2(\theta_2 - \phi)}
                       {1 + \frac{\sigma^2_{2,\rm maj}}{\sigma^2_{2,\rm min}} \tan^2(\theta_2 - \phi)} \, \sigma^2_{2,\rm maj},
\eeq{e:e2}
\end{widetext}
   where $\phi = \arctan{\Frac{\Delta\delta}{\Delta\alpha\cos\delta}}$.

  Second, I computed the histograms of the ratios of arc lengths between
positions estimates to the arc length uncertainties --- the so-called 
normalized are lengths. If position uncertainties are correct, 
the distribution of normalized arc lengths should be Rayleighian with 
$\sigma=1$. Third, I performed a multiplicative fitting of the 
uncertainties to the Rayleigh distribution. I considered that the position 
uncertainties from both odd and even decimation solutions are to be scaled 
by a common factor $R$ and sought such a factor that minimizes the sum of 
squares of the differences of the normalized histogram and the Rayleigh 
distribution. When $R>1$, the errors are underestimated. When $R<1$, 
the errors are overestimated. I have rescaled position uncertainties by 
dividing them by $\sqrt{R}$ in further analysis.

\subsection{Quasi-simultaneous observations at 2.2/8.4~GHz and 
            4.1/7.4~GHz with VLBA}

\begin{figure}
   \includegraphics[width=0.49\textwidth]{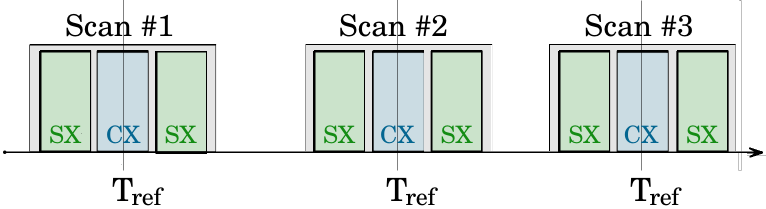}
   \caption{Data were recorded three times for every scan during BP175 campaign
            when antennas were on source: first at SX, then at at CX, and 
            then again at SX.}
   \label{f:bp175}
\end{figure}

   In order to assess the impact of a frequency change of dual-band
observations on source position estimates, a special 48~hr campaign BP175 
was observed with VLBA. The campaign was split into ten blocks of 3 to 
8~hours long. Data from each pointing was recorded three times: first at 
2.2/8.4~GHz (SX bands), then the receiver was changed to 4.1/7.4~GHz (CX), 
and then back to SX (see Figure~\ref{f:bp175}). More detail of this campaign 
can be found in \citep{r:wfcs}. In total, 13,512 group delays at all bands 
were computed at exactly the same reference epoch. SX and CX group delays were 
processed independently, and positions of 394 observed sources were estimated
in two separate least square (LSQ) solutions. 

\subsection{The geodesy and absolute astrometry VLBI campaign at 2.2/8.4~GHz 
            with VLBA}

\begin{figure}
   \includegraphics[width=0.49\textwidth]{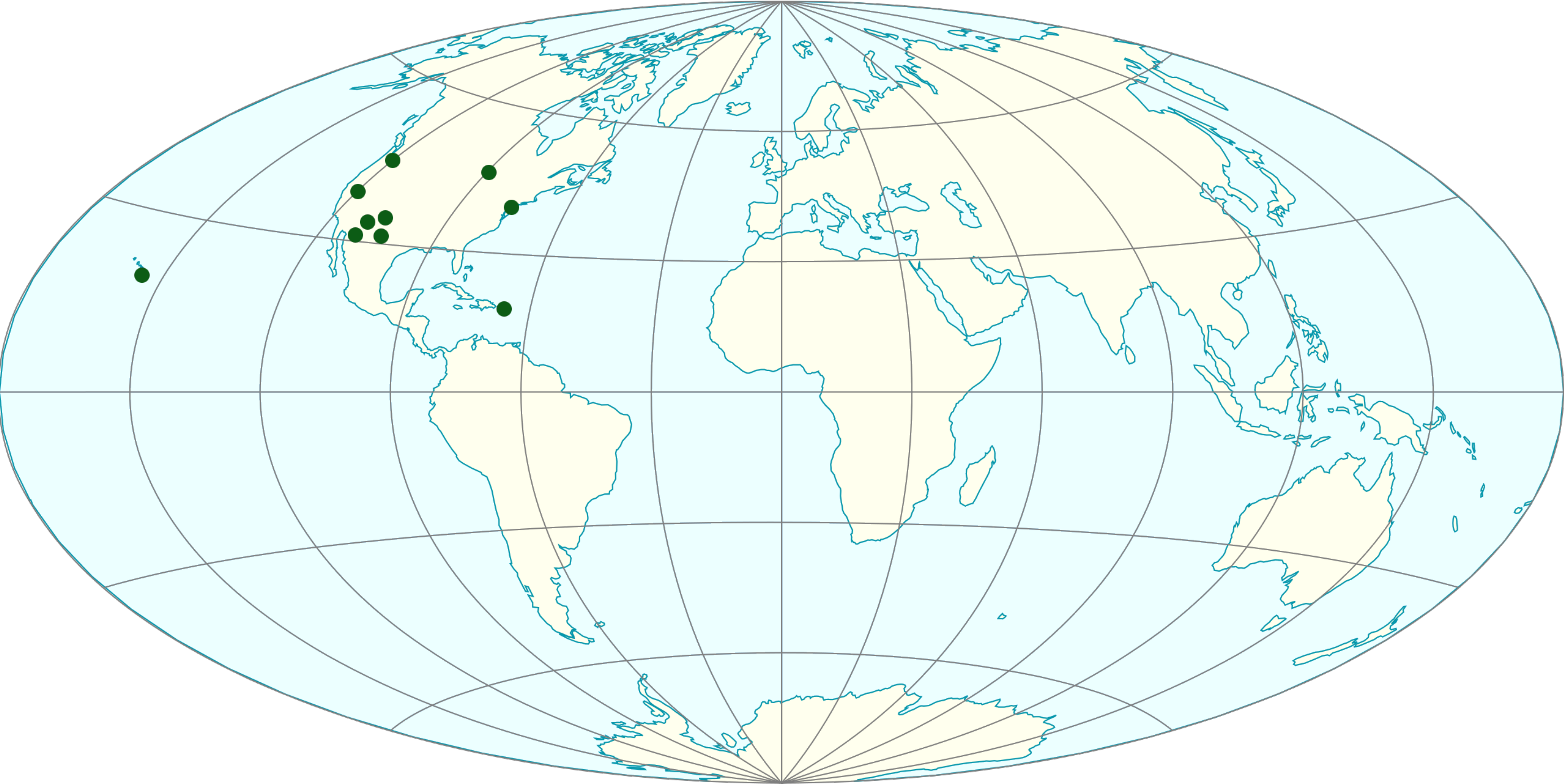}
   \caption{Location of 10 VLBA radiotelescopes.}
   \label{f:vlba_map}
\end{figure}

  A program of regular twenty four hour observations with VLBA and up to 10 
other stations runs since 1994 through present with a cadence of 
approximately 6 experiments per year. See the distribution of VLBA stations 
in Figure~\ref{f:vlba_map}. The purpose of that program is monitoring positions 
of radiotelescopes and improvement of source coordinates. Observations are done 
at 2.2 and 8.4~GHz simultaneously with a spanned bandwidth of 140 and 496~MHz 
respectively. More information about this campaign can be found in  
\citet{r:rdv}. I included in data analysis experiments from similar geodetic 
programs cn18 and cn19 and three campaigns of the second epochs of VLBA 
Calibrator Survey: bg219, ug002, and ug003 \citep{r:vcs-ii}. I selected 
270 experiments since April 15 1998 through November 28 2022, in total 
3.4~million observations. Data from stations other than VLBA were discarded 
in the solutions in order to eliminate the impact of the network for 
comparison with other datasets that used only VLBA antennas. The R-factor 
from the decimation solutions is 0.94.
  
\subsection{The absolute astrometry VLBI campaign at 4.1/7.4~GHz with VLBA}

  A program of the wide-field survey with VLBA at 4.1/7.4~GHz ran in 
2013--2022 \citep{r:wfcs}. That program targeted 29,851 sources. Most of the 
target sources were weak with correlated flux densities in a range of 
10--100~mJy, and only one half have been detected. In addition to target 
sources, a number of strong sources were observed as calibrators. It is just
observations of these calibrator sources that were mainly used in 
comparisons. I included data of 281 experiments, 0.6 million observations, 
under this program for data analysis. The R-factor from the decimation 
solutions is 1.24.

\subsection{The geodesy VLBI campaign at 2.2/8.4~GHz at the IVS network}
  
\begin{figure}[h]
   \includegraphics[width=0.49\textwidth]{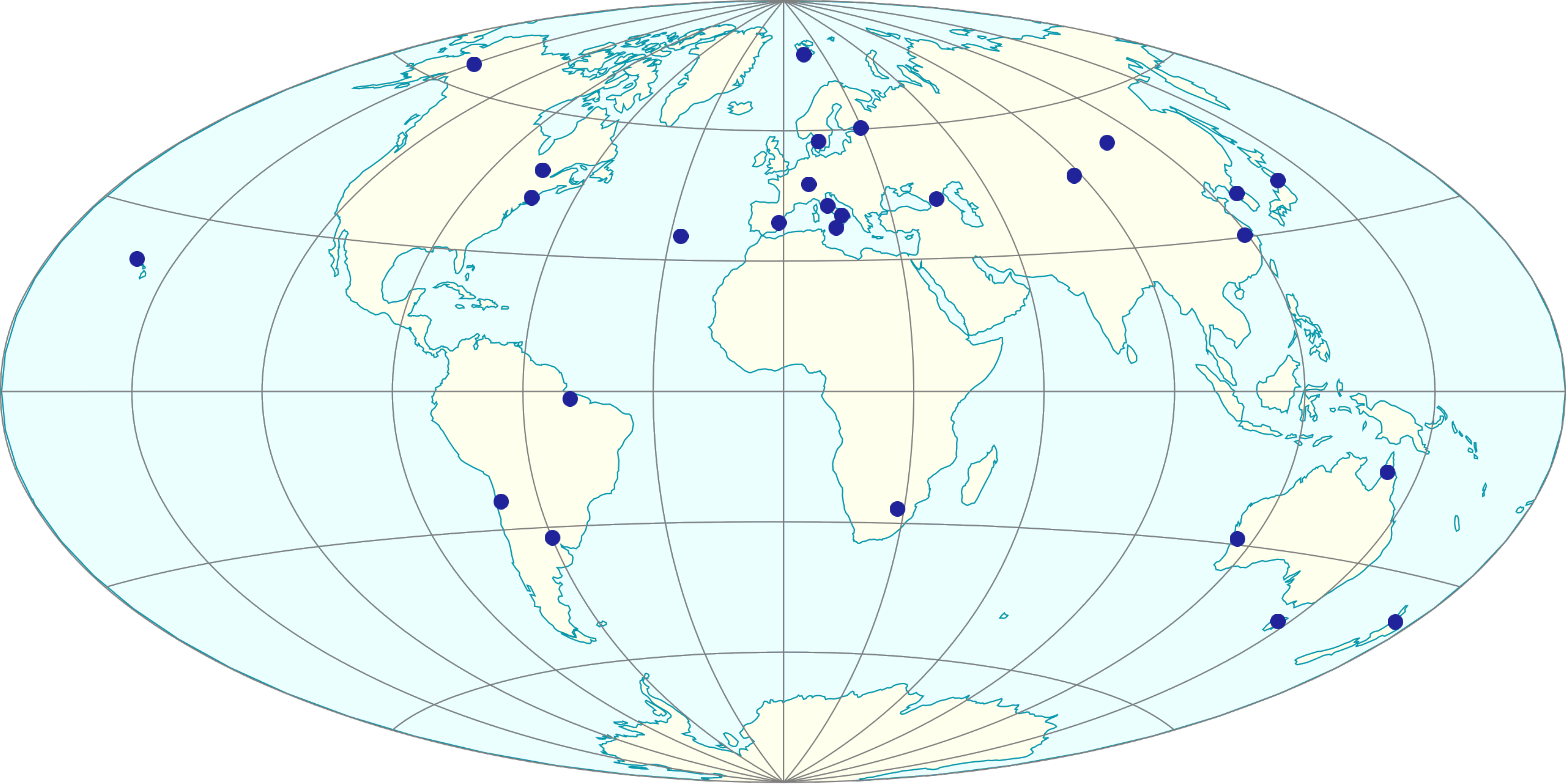}
   \caption{Location of 33 radiotelescopes that participated in IVS SX 
            R1R4 campaigns.}
   \label{f:r1r4_map}
\end{figure}

   An on-going campaign of regular geodetic VLBI observations R1 and 
R4 \citep{r:r1r4} runs two times a week on Mondays and Thursdays since 2002.
There are 43 stations that joined these observations. I retained 33 stations 
that participated in 30 or more experiments for data analysis. See station
distribution in Figure~\ref{f:r1r4_map}. In total, 6.9 million group delays 
from 2259 twenty-four hour experiments were used in data analysis. 
The R-factor from the decimation solutions is 1.13.

\subsection{The geodesy VLBI campaign VO at 3.0/5.2/6.4/10.2~GHz at the IVS network}

\begin{figure}[h]
   \includegraphics[width=0.49\textwidth]{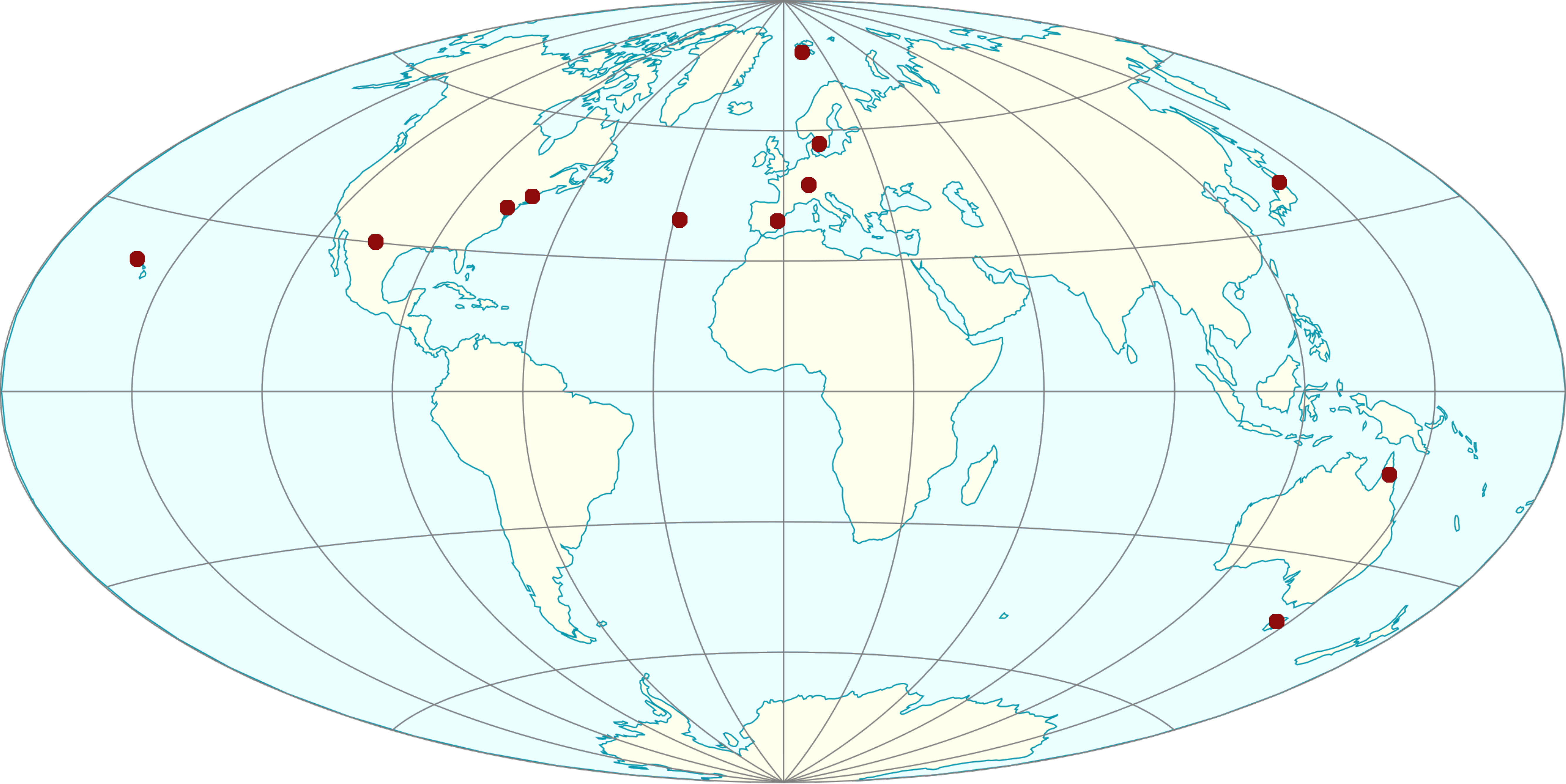}
   \caption{Location of 13 radiotelescopes that participated in the IVS
            quad-band campaign.}
   \label{f:vgos_map}
\end{figure}

   Another ongoing campaign of regular geodetic VLBI observations VO
\citep{r:vgos} runs at a network of 13 twelve meter radiotelescopes 2--4 
times a month since 2017. The network of stations evolved with time. 
The map of station distribution by the end of 2023 is shown in 
Figure~\ref{f:vgos_map}. Observations are performed simultaneously at four 
bands 3.0, 5.2, 6.4 and 10.2~GHz. Ionosphere-free group delays are directly 
estimated from these data during fringe fitting. In total, 1.3~million 
group delays from 153 experiments were used in data analysis. The R-factor 
from the decimation solutions is 1.53.
  
\subsection{The absolute K~band astrometry VLBI campaign at 23~GHz with VLBA}

  A number of absolute astrometry campaigns ran with VLBA at K~band 
(23--24~GHz) since 2002. They include VCS5 campaign \citep{r:vcs5}, KQ-survey 
\citep{r:kq1,r:kq2}, Kband-CRF \citep{r:witt23}, BJ083, UD001, UD009 
\citep{r:kba}, and UD015 campaigns. The recorded bandwidth varied from 32~MHz 
at single right circular polarization in early experiments to 512~MHz, dual 
polarization in 19 latest observing sessions. In total, 99 experiments, 
1.5 million group delays were used in data analysis. Of 1298 observed 
radio sources, 1126 have been detected. The R-factor from the decimation 
solutions is 0.92.

\subsection{The absolute Q-band astrometry VLBI campaign at 43~GHz with VLBA}

  There were two VLBA absolute astrometry observing campaigns at Q-band 
(43~GHz) with VLBA: KQ-survey in 2002--2003 and UD014 in 2021. These campaigns 
targeted mainly the strongest sources. In total, 72,498 group delays from 
8 experiments were used in data analysis. Of 525 observed sources, 504 have 
been detected. The R-factor from the decimation solutions is 1.17.

\section{Analysis of the differences in source  position derived from seven datasets}

   I estimated source positions using each dataset. Then I formed the 
differences in positions of those sources that were observed and have been 
detected in both campaigns and which positions were derived with 
a sufficient accuracy in order to investigate small differences. I used two 
criteria to select common sources for the statistical study: the number of 
observations used in solutions and the position uncertainty. I normalized 
residuals by dividing them by reweighted uncertainties. In order to mitigate 
the impact of outliers on the source statistics, I performed an iterative 
procedure of outlier elimination. I retained for further analysis only 
the sources with normalized residuals by module less than some number, 
by default 5.

\subsection{Comparison of source positions at 4.1/7.4~GHz vs 2.2/8.4~GHz}

  Comparison of quasi-simultaneous 2.2/8.4 versus 4.1/7.4~GHz did not reveal 
a systematic a pattern. The rms difference was 0.54~mas over declination and 
0.56~mas over right ascension scaled by $\cos\delta$. It should be noted that 
that campaign was a pilot study of a large observing program focused on 
observations of weak sources. This explains relatively large position 
uncertainties and differences in position estimates.

\begin{figure}[h]
   \includegraphics[width=0.49\textwidth]{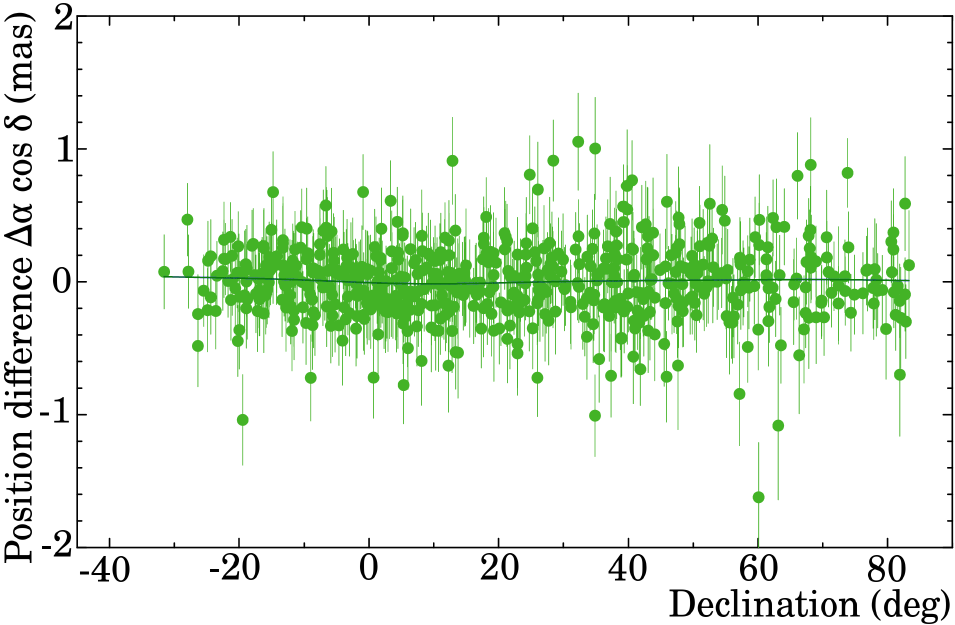}
   \caption{The differences in right ascension scaled by $\cos\delta$ factor
            derived from analysis of 2.2/8.4 and 4.1/7.4~GHz VLBA 
            observations.
           }
   \label{f:xsxc_ra}
\end{figure}

\begin{figure}[h]
   \includegraphics[width=0.49\textwidth]{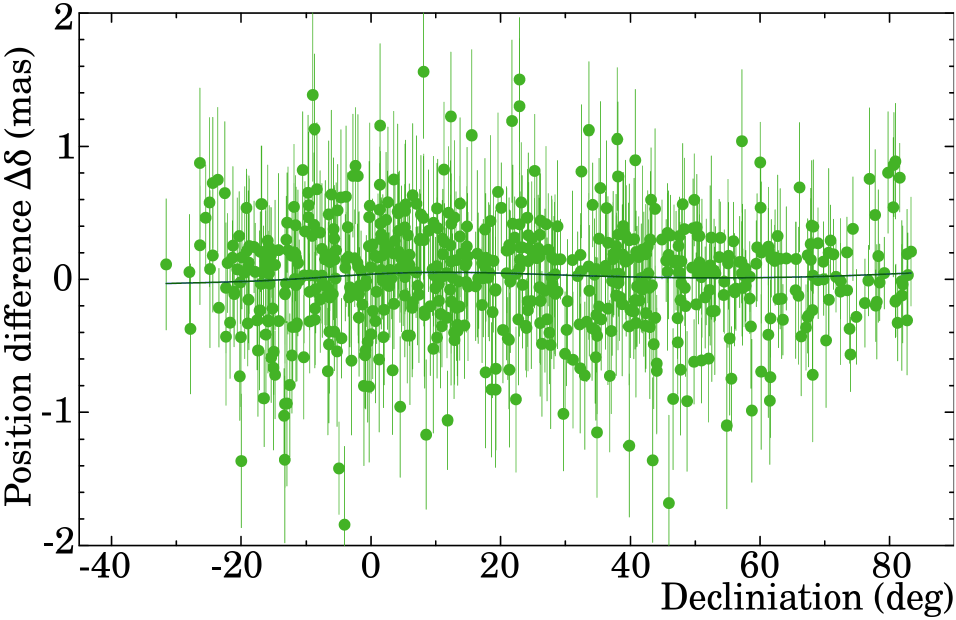}
   \caption{The differences in declination derived from analysis of 
            2.2/8.4 and 4.1/7.4~GHz VLBA observations. 
           }
   \label{f:xsxc_dec}
\end{figure}

  Processing a significantly larger dataset of non-simultaneous 2.2/8.4~GHz
and 4.1/7.4~GHz observations from the VLBA network allowed me to lower the 
limit of position differences. Plots of differences are featherless --- see 
Figures~\ref{f:xsxc_ra}--\ref{f:xsxc_dec}; the biases in right ascension
and declination are 0.004 and 0.028~mas respectively, and the rms of the 
differences is 0.27 and 0.39~mas over right ascension and declination
among 734 common sources with formal uncertainties greater than 0.5~mas.

\begin{figure}[h]
   \includegraphics[width=0.49\textwidth]{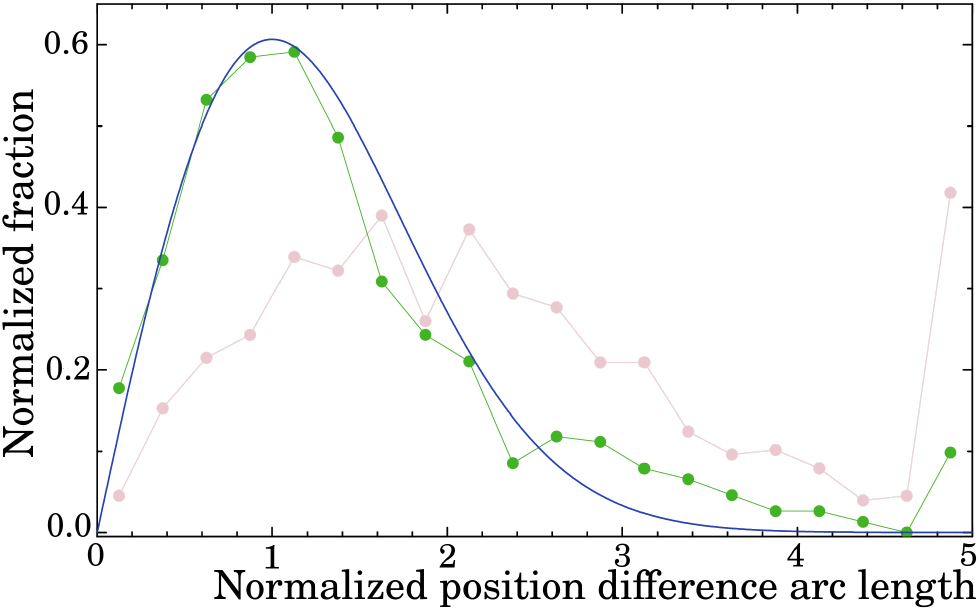}
   \caption{The differences in normalized arc lengths between source 
            source positions derived from analysis of 2.2/8.4 and 
            4.1/7.4~GHz VLBA observations. Pale pink points show the 
            histogram computed with original uncertainties and green 
            points show the show the histogram computed with inflated 
            uncertainties.
           }
   \label{f:xsxc_arc}
\end{figure}

  The histogram of normalized arc lengths between position estimates 
computed using the original uncertainties displays a significant 
deviation from the Rayleigh distribution (Figure~\ref{f:xsxc_arc}). 
I computed a series of histograms of normalized differences in right 
ascension scaled by $\cos\delta$ and declination with different variances 
that were added in quadrature to position uncertainties. I found the 
variances that provided the minimum of the rms differences between 
a histogram and the Gaussian distributions. These variances are 0.11 and 
0.09~mas over right ascension and declination respectively. The 
distribution of the normalized arc lengths became much closer to the 
Rayleigh distribution after inflating the uncertainties.

\subsection{Comparison of source positions from quad-band observations 
            versus positions from 2.2/8.4~GHz IVS observations.}

\begin{figure}[h]
   \includegraphics[width=0.49\textwidth]{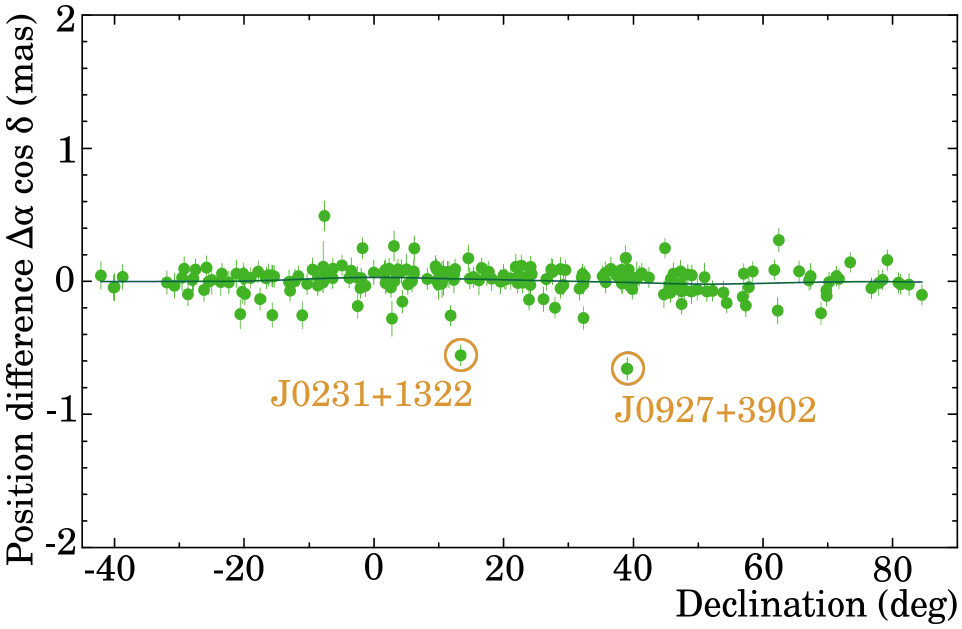}
   \caption{The differences in right ascension scaled by $\cos\delta$ factor
            derived from analysis of 3.0/5.2/6.4/10.2 and 2.2/8.4~GHz 
            geodetic VLBI observations.
           }
   \label{f:vgos_r1r4_ra}
\end{figure}

\begin{figure}[h]
   \includegraphics[width=0.49\textwidth]{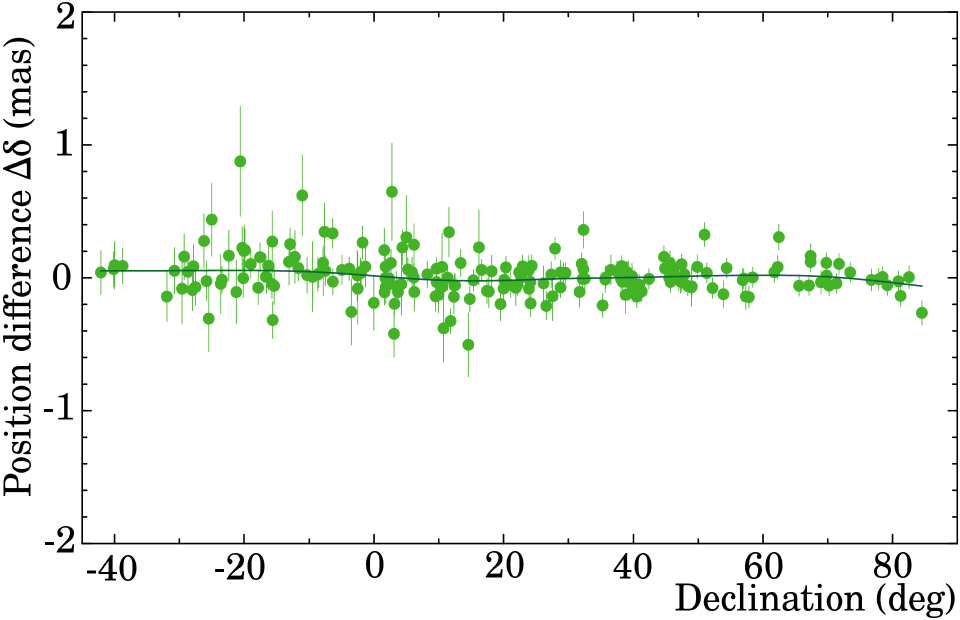}
   \caption{The differences in declination derived from analysis 
            of 3.0/5.2/6.4/10.2 and 2.2/8.4~GHz 
            geodetic VLBI observations.
           }
   \label{f:vgos_r1r4_dec}
\end{figure}

  VLBI observing program at 3.0/5.2/6.4/10.2 GHz with the so-called
VLBI Global Observing System (VGOS) network dedicated to determination of 
the Earth orientation parameters and station positions intensively used 
a short list of sources. There are 204 sources with 32 or more good 
observations common among other programs. Plots of the differences are shown 
in Figures~\ref{f:vgos_r1r4_ra}--\ref{f:vgos_r1r4_dec}. There are two 
outliers in right ascension differences, but otherwise, the plots do not 
exhibit systematic patterns.

  In a similar way, as we saw in the prior comparison, the histogram of 
normalized arc lengths does not agree with the Rayleigh distribution 
despite of scaling VGOS position uncertainties by the factor of 1.53 derived
from the decimation test. I found that the uncertainties over right 
ascension and declination should be inflated by adding in quadrature
0.070 and 0.086~mas respectively to fit their normalized histogram
to the Gaussian distribution. Biases in right ascension and declination
are 0.006 and 0.007~mas respectively. The rms of position differences
over right ascension and declination are 0.09 and 0.13~mas respectively.

\begin{figure}[h]
   \includegraphics[width=0.49\textwidth]{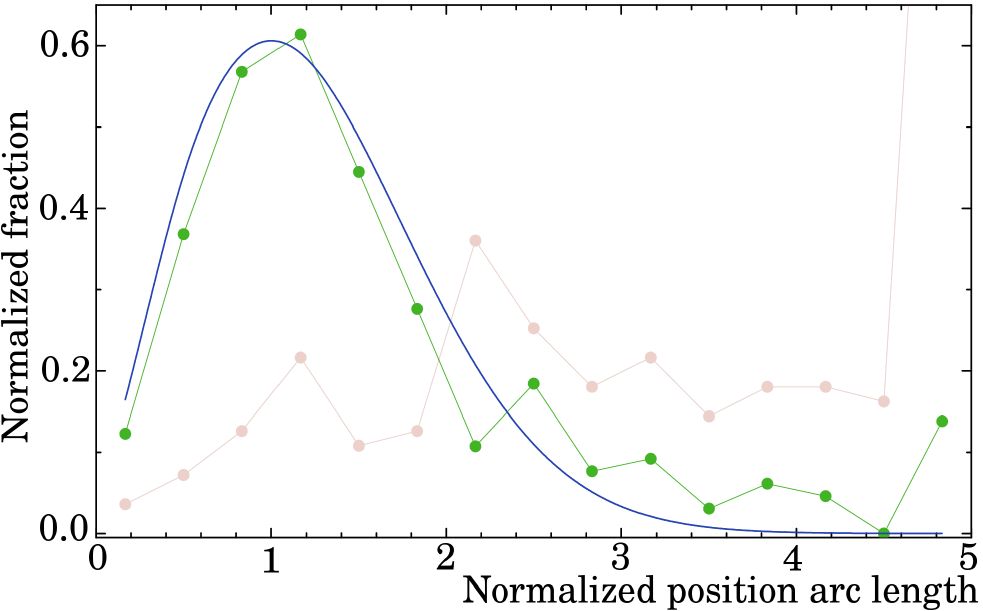}
   \caption{The differences in normalized arc lengths between position
            estimates derived from analysis of 3.0/5.2/6.4/10.2 and 
            2.2/8.4~GHz geodetic VLBI observations. Pale pink points show 
            the histogram computed with original uncertainties and green 
            points show the show the histogram computed with inflated 
            uncertainties.
           }
   \label{f:vgos_r1r4_arc}
\end{figure}

\subsection{Comparison of source positions from 2.2/8.4~GHz at the IVS 
            and VLBA networks.}

\begin{figure}[h]
   \includegraphics[width=0.49\textwidth]{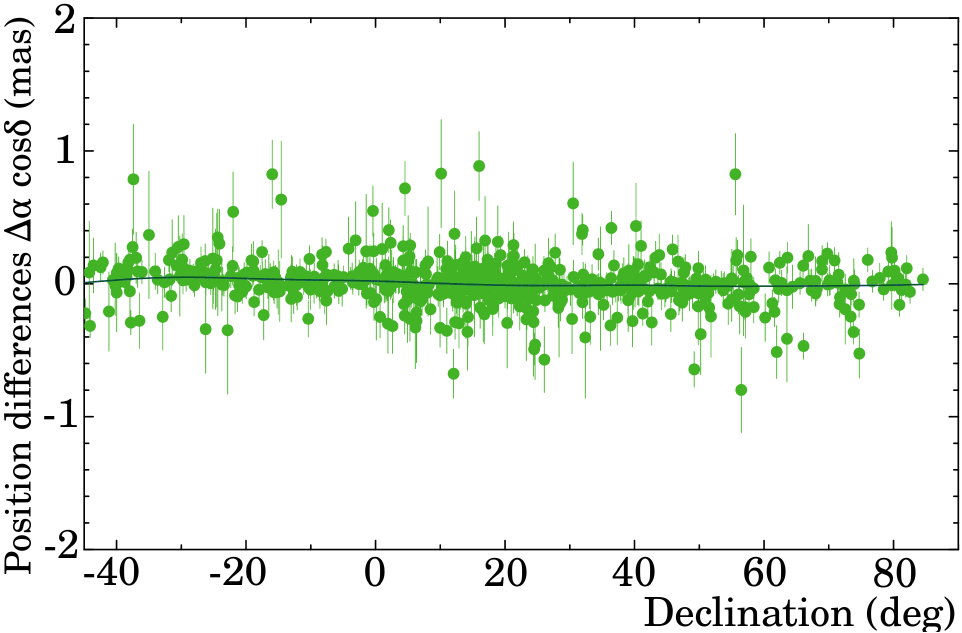}
   \caption{The differences in right ascension scaled by $\cos\delta$ factor
            derived from analysis of 2.2/8.4~GHz observations at VLBA 
            and R1R4 networks.
           }
   \label{f:xs_r1r4_ra}
\end{figure}

\begin{figure}[h]
   \includegraphics[width=0.49\textwidth]{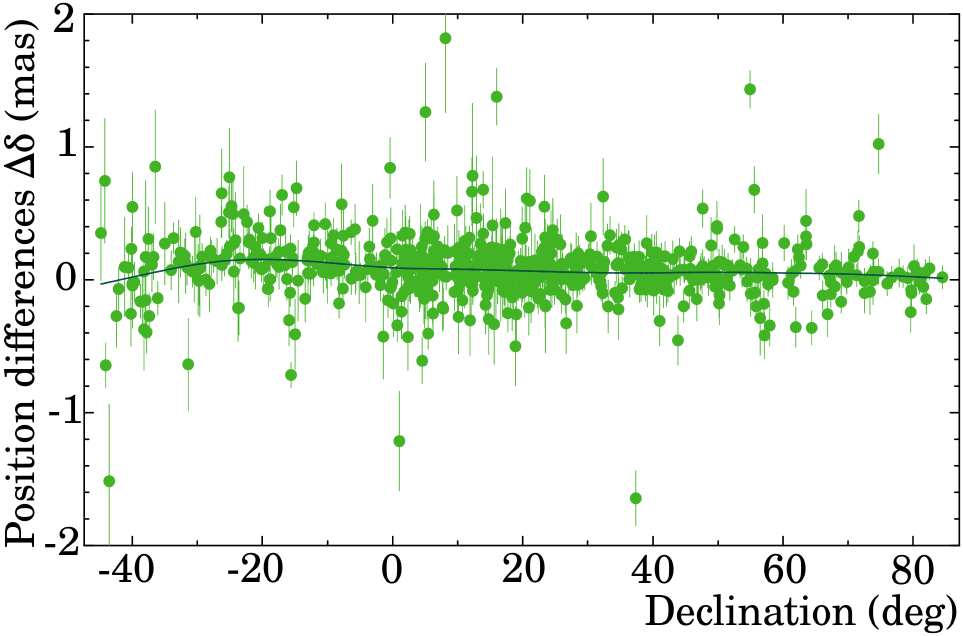}
   \caption{The differences in declination derived from analysis 
            of 2.2/8.4~GHz observations at VLBA and R1R4 networks.
           }
   \label{f:xs_r1r4_dec}
\end{figure}

  It is instructive to extend analysis of position differences determined 
from observations at the same network and different frequencies, to 
the differences derived from observation at different networks and at 
the same frequencies. This analysis allows us to assess the magnitude of 
frequency dependent systematic errors with respect to frequency 
independent errors. Plots of differences in position estimates of 665 
common sources with formal uncertainties less than 0.5 mas and at least 
32 good observations are shown in 
Figure~\ref{f:xs_r1r4_ra}--\ref{f:xs_r1r4_dec}. We see a number of 
outliers and a small declination bias at low declinations. The overall 
bias in right ascension is 0.004~mas and 0.074~mas in declination. 

  Fitting the normalized position differences to the Gaussian distribution, 
I found that the uncertainties over right ascension and declination should 
be inflated by adding in quadrature 0.060 and 0.087~mas respectively to 
provide the best fit. These parameters are remarkably close to those
found in the comparison between VGOS and VLBA observations.  The rms of
the differences in right ascension and declination are 0.090 and 0.138~mas 
respectively. 

\begin{figure}[h]
   \includegraphics[width=0.49\textwidth]{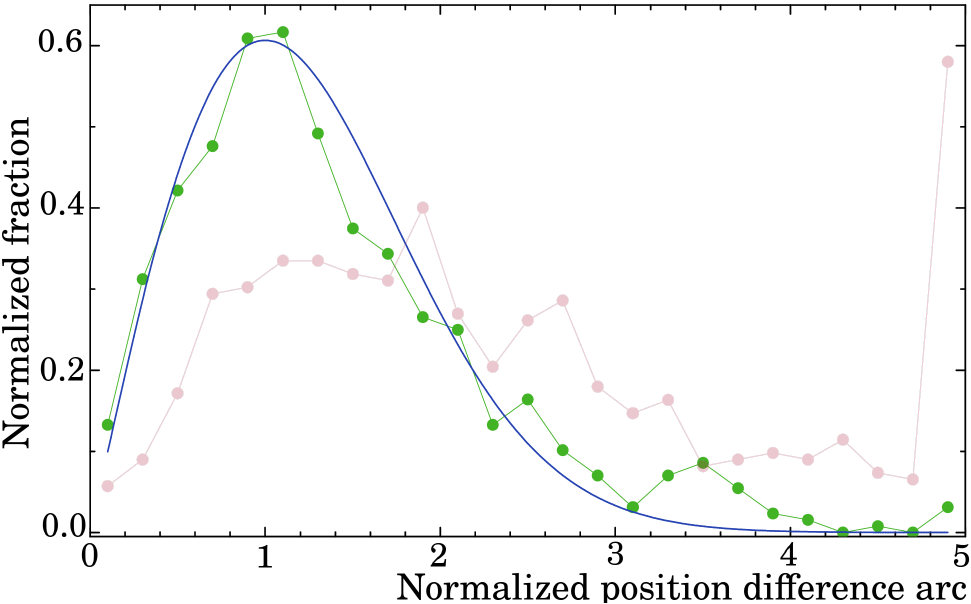}
   \caption{The differences in normalized arc lengths between source 
            positions derived from analysis of 2.2/8.4~GHz data from 
            VLBA and R1R4 networks. Pale pink points show the histogram 
            computed with original uncertainties and green points show the 
            the histogram computed with inflated uncertainties.
           }
   \label{f:xs_r1r4_arc}
\end{figure}

  For completeness, I analyzed the differences between source position
estimates derived from VGOS and VLBA SX data. The declination bias was 
-0.056~mas, the extra noise in right ascension and declination was 0.071
and 0.064~mas respectively. The rms of the differences in right ascensions
and declination are 0.098 and 0.138~mas respectively. This triple 
comparison helped to identify the source catalogue that has a declination 
bias with respect to other catalogues: VLBA SX.

  When we consider the difference between two source position catalogue,
in general, we cannot tell what is the contribution to these differences 
of each individual catalogue. However, when differences of {\it three} 
catalogues of the same sources are available, we can do it using the 
so-called three corner hat method \citep{r:tri_corner_hat}. Considering 
that the intrinsic errors in catalogues $c_1, c_2, c_3$ do not have 
correlations between each others, we can write
\beq
   \begin{array}{lcl}
       \Var(c_1 - c_2) & = & \Var(c_1)  + \Var(c_2)  \\
       \Var(c_1 - c_3) & = & \Var(c_1)  + \Var(c_3)  \\
       \Var(c_2 - c_3) & = & \Var(c_2)  + \Var(c_3)  \\
   \end{array}.
\eeq{e:e3}

  Solving equation~\ref{e:e3}, we get
\begin{widetext}
\beq
   \begin{array}{lcl}
       \Var(c_1) = \Frac{1}{2}( \Var(c_1 - c_2) + \Var(c_1 - c_3) - \Var(c_2 - c_3) ) \\
       \Var(c_2) = \Frac{1}{2}( \Var(c_2 - c_3) + \Var(c_1 - c_2) - \Var(c_1 - c_3) ) \\
       \Var(c_3) = \Frac{1}{2}( \Var(c_1 - c_3) + \Var(c_2 - c_3) - \Var(c_1 - c_2) ) \\
   \end{array}.
\eeq{e:e4}
\end{widetext}

 This method allows us to solve for intrinsic errors of each catalogue. 
Here I used 197 sources common in all three catalogues. The result is 
presented in table~\ref{t:cat_err}. We see that the intrinsic errors of 
individual catalogues are at a level of 0.05--0.07~mas in right ascension 
and declination, except the SX VLBA catalogue that has intrinsic errors 
in right ascension at a level of 0.11~mas. The median formal uncertainties 
scaled by a factor of $R$ from the decimation test are shown as well 
for comparison.

\begin{table}
   \caption{The intrinsic (columns 2 and 3) and formal errors (columns 4 and 5)
            of source positions catalogues from VLBA, VGOS, and 
            R1R4 catalogues in mas. $\cos\delta$ factor is applied to errors
            in right ascension.
           }
   \makeatletter\if@two@col \hspace{-4em} \fi\makeatother
   \begin{tabular}{lrrrr}
       \hline
               &  $\sigma_\alpha$(i) & $\sigma_\delta$(i) & 
                  $\sigma_\alpha$(f) & $\sigma_\delta$(f)  \\
       \hline
        R1R4 SX        & 0.05 & 0.06 & 0.01 & 0.03 \\
        VGOS quad-band & 0.10 & 0.05 & 0.02 & 0.04 \\
        VLBA SX        & 0.05 & 0.07 & 0.02 & 0.03 \\
       \hline
    \end{tabular}
    \label{t:cat_err}
\end{table}

\subsection{Comparison of K~band positions against SX~positions}

  There are in total 1007 common radio sources that have been detected in both 
VLBA datasets at 23~GHz and 2/8~GHz. I used in further analysis 858 sources 
that had at least 32 common observations in both datasets, position 
uncertainties $< 0.5$~mas over declination or right ascension scaled by
$\cos\delta$, and normalized residuals less than 7.
  
\begin{figure}[h]
   \includegraphics[width=0.49\textwidth]{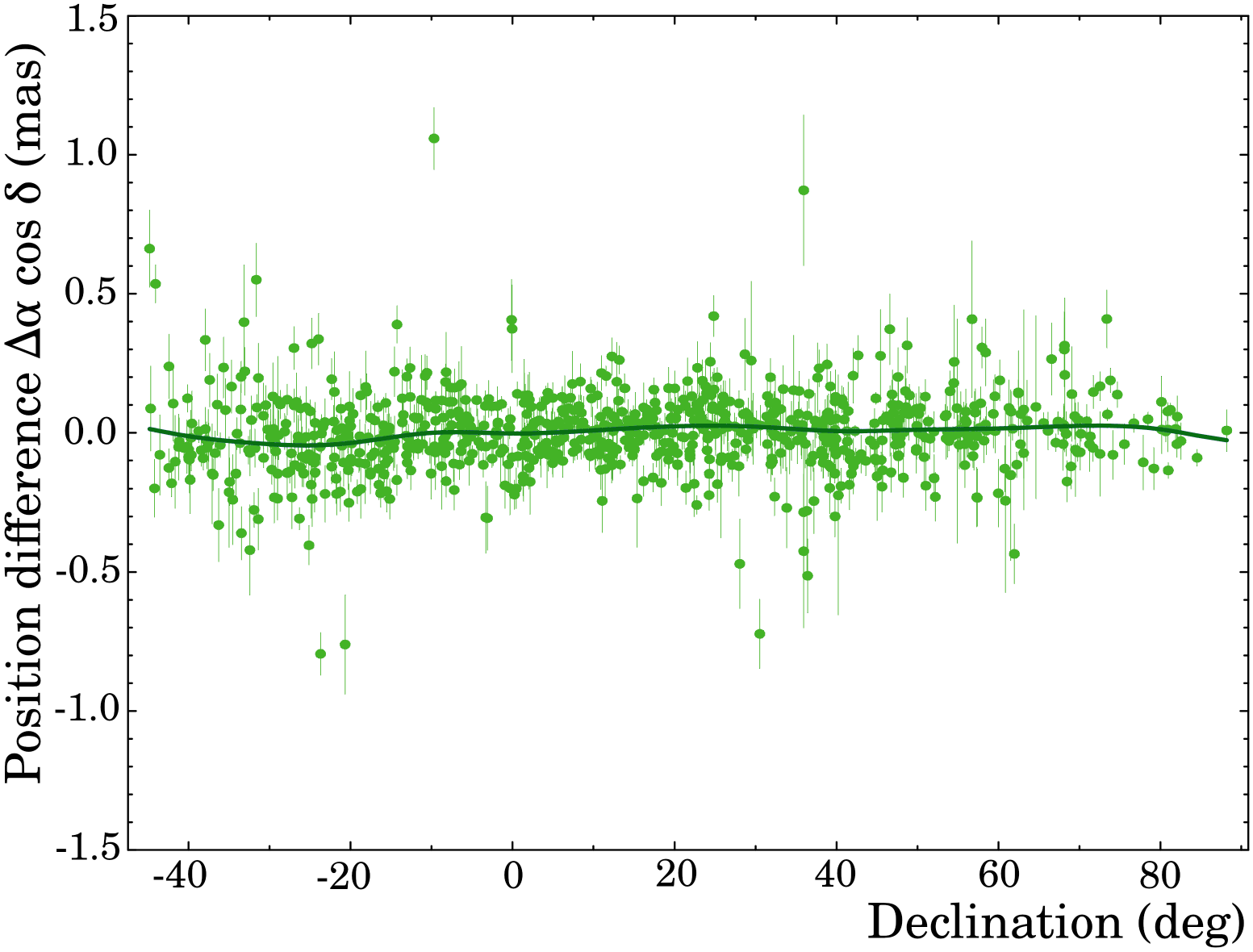}
   \caption{The differences in right ascension scaled by $\cos\delta$ from 
            analysis of VLBA data at 2.2/8.4 GHz versus 23~GHz.}
   \label{f:k_xs_ra}
\end{figure}

\begin{figure}[h]
   \includegraphics[width=0.49\textwidth]{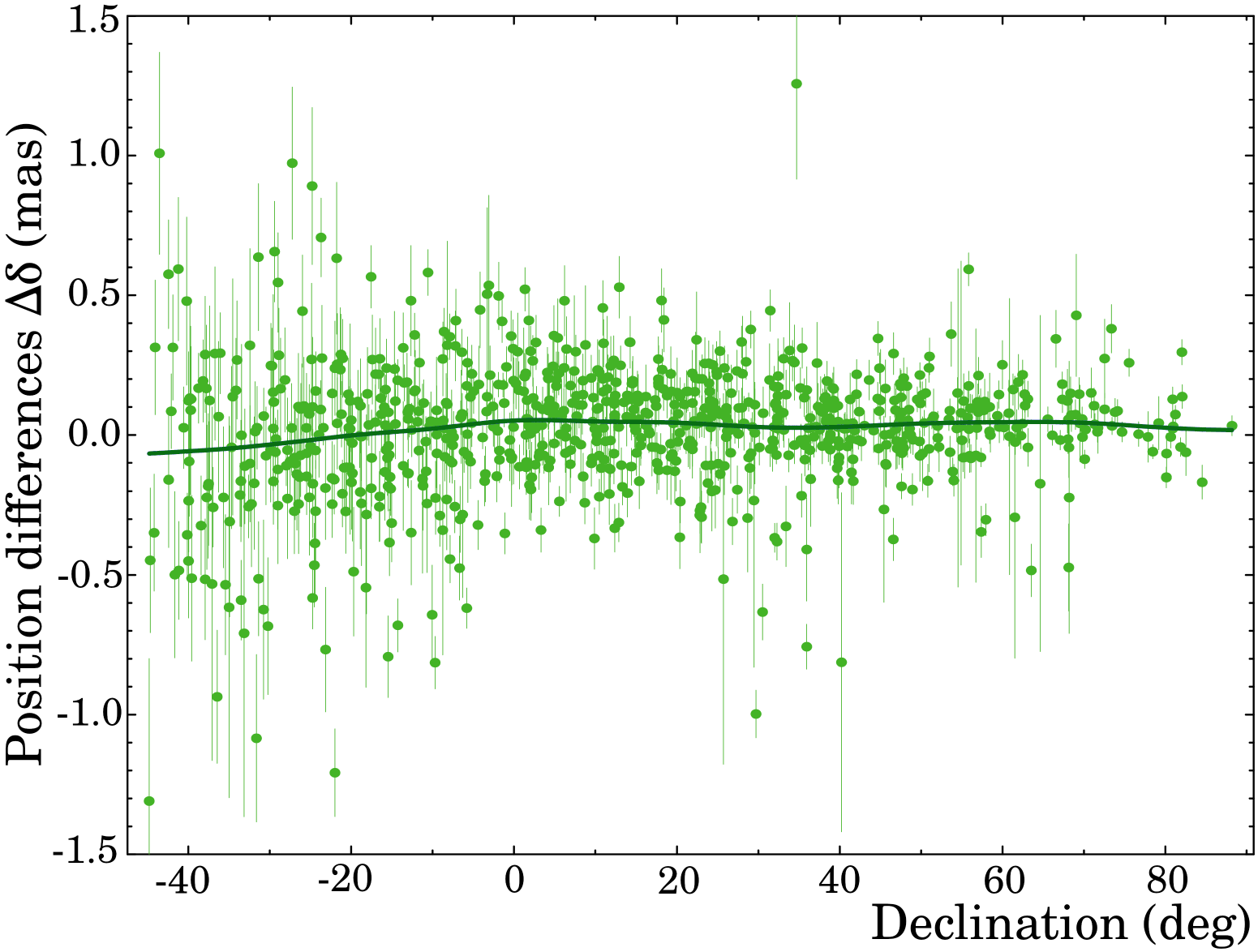}
   \caption{the differences in declination from 
            analysis of VLBA data at 2.2/8.4 GHz versus 23~GHz.}
   \label{f:k_xs_dec}
\end{figure}

  The position differences in right ascension scaled by $\cos\delta$ factors 
as a function of declination are shown in 
Figures~\ref{f:k_xs_ra}--\ref{f:k_xs_dec}. The solid lines are the boxcar 
averages. Table~\ref{t:k_xs} summarizes the statistics of the sample. While 
the plot of differences in right ascension is featureless, except some 
outliers, the plot of differences in declination shows a systematic pattern: 
the declination uncertainties of the sources in the Southern Hemisphere are 
noticeably larger and their scatter is greater. 

\begin{table}
   \caption{Statistics of positions uncertainties of source position
            from SX and K band VLBA observations in mas.}
   \begin{tabular}{lrrr}
       \hline
                                &  SX   & K      & Diff  \\
       \hline
        RA $\cos\delta$ median  & 0.051 & 0.036  &       \\
        Dec median              & 0.085 & 0.061  &       \\
        RA $\cos\delta$ bias    &       &        & 0.002 \\
        RA $\cos\delta$ rms     &       &        & 0.109 \\
        Dec bias                &       &        & 0.033 \\
        Dec rms                 &       &        & 0.156 \\
        Arc median              &       &        & 0.152 \\
       \hline
    \end{tabular}
    \label{t:k_xs}
\end{table}

  Observations of southern sources with the VLBA array situated in the
Northern Hemisphere are made in a disadvantageous configuration compared
with observations of northern sources. These sources usually cannot be 
seen at all the baselines, and the lower declination of a given source, 
the less stations can simultaneously see it. Since two the southmost 
VLBA stations, {\sc mk-vlba} in Big Hawaiian island and {\sc sc-vlba} in 
St.~Croix island in the Caribbean sea have also the widest spread in 
longitude, southern sources are often observed at a sub-network with 
a long equatorial baseline vector projection and a short polar baseline 
vector projection. This explains a disparity in position uncertainties 
between declination and right ascension. However, a more detailed analysis 
reveals that the declination dependence of differences in declination
is stronger than the declination dependence of uncertainties. This 
indicates there is another factor that affects the differences beyond 
a purely geometric effect.

\begin{figure}[h]
   \includegraphics[width=0.49\textwidth]{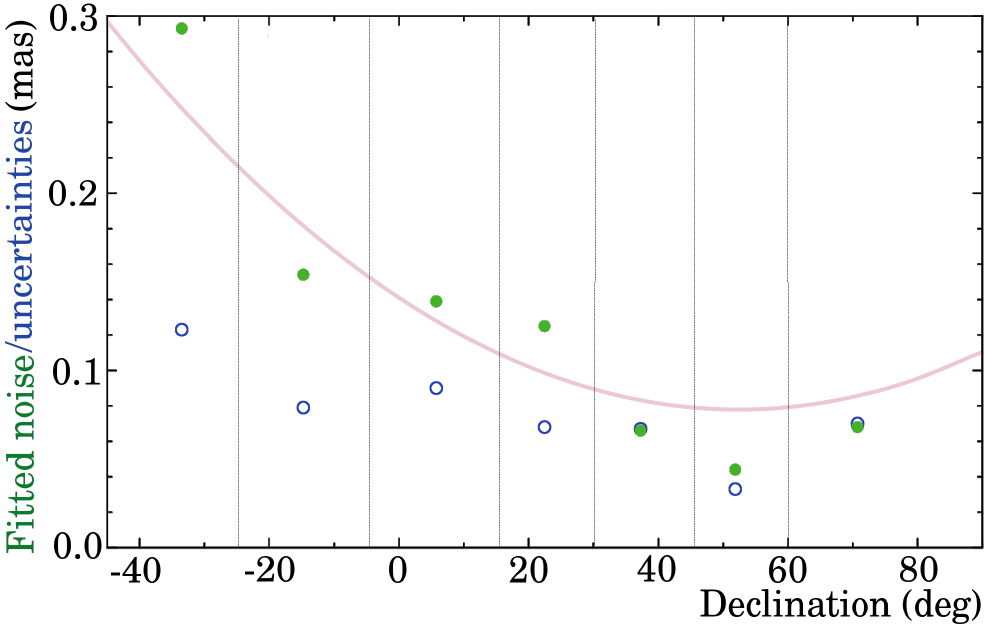}
   \caption{Green solid circles show position differences in declination from 
            analysis of VLBA data at 2.2/8.4 GHz versus 23~GHz at
            7~declination segments. Blue hollow circles show the median
            K~band position uncertainty. The pink solid line shows a predicted
            additional variances in declination due to residual errors in
            the ionospheric contribution to path delay \citep{r:sba}.
           }
   \label{f:k_xs_dec_seg}
\end{figure}

  I split the dataset into seven segments over declinations and computed
statistics within each segment. In Figure~\ref{f:k_xs_dec_seg} blue hollow
circles show the median 23~GHz position uncertainties and green solid 
circles show median differences in declination. Both statistics grow
with declination, but the position differences grow faster with a decrease
in declination. \citet{r:sba} investigated the impact of residual 
ionospheric errors on source positions after applying the contributions 
derived from the GNSS global ionospheric model. I showed in that paper that 
the residual contribution causes a declination dependent extra noise in source 
positions derived from processing data at VLBA. The origin of this declination 
dependence is the latitude 
dependence of the electron contents in the ionosphere: the total electron 
contents in the equatorial bulge is up to one order of magnitude higher than 
in the polar regions. Since observations of low declination sources at the 
northern arrays such as VLBA have to be done mainly in the southern directions 
where the total electron contents is systematically higher, the residual 
ionospheric path delay contribution is systematically higher with respect 
to observations of high declination sources. The pink thick line in 
Figure~\ref{f:k_xs_dec_seg} shows the K~band extra variance in 
declination (Figure~14 in \citet{r:sba}) derived from the comparison of 
the ionospheric path delay from dual-band SX VLBA observations and path
delay from the GNSS ionospheric model and then scaled by the square of 
the frequency ratio $(23.7/8.6)^2 = 7.6$. We see that the observed
growth of differences in declination with declination is in a reasonable
agreement with that model. It should be stressed that the model of the 
increased errors in declination was derived without any knowledge of 
K~band astrometry, and therefore, can be considered as an independent 
source of information.

\begin{figure}[h]
   \includegraphics[width=0.49\textwidth]{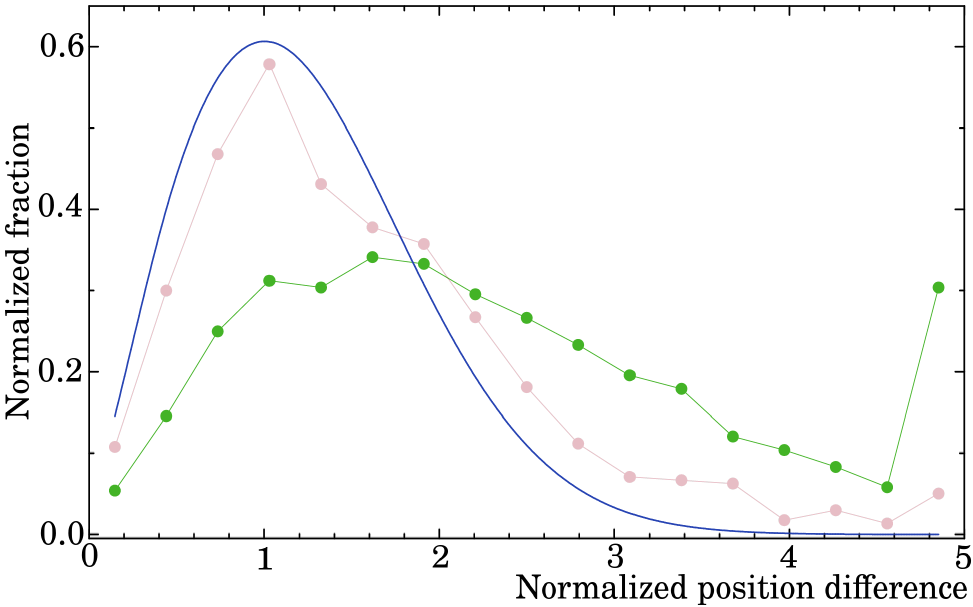}
   \caption{The distribution of normalized arc lengths between K~band and SX
            position estimates. Green bright circles show the distribution
            using original data. Pale pink circles show the distribution
            with the contribution of the residual ionospheric noise applied.
            The thick blue line shows the Rayleigh distribution with 
            $\sigma=1$ as a reference.
           }
   \label{f:k_xs_raw}
\end{figure}

   Inspired by the agreement of the model with observations, I decided to 
investigate the differences in astrometric positions further and attempted
to solve an ambitious problem: to build a {\it quantitative} stochastic model 
of the differences. Figure~\ref{f:k_xs_raw} shows the normalized histogram
of normalized arc lengths between K~band and SX source position estimates. If 
there were no systematic errors, and the reported uncertainties were correct, 
the distribution should have been Rayleighian with $\sigma=1$ (blue thick 
line in Figure~\ref{f:k_xs_raw}). We see significant discrepancies. Then
I added in quadrature the extra noise in source positions due to mismodeling
the ionospheric contribution to group delay shown in Figure~14 of 
\citet{r:sba} to the K~band source position uncertainties. Specifically, I used 
the following regression:
\beq
  \begin{array}{lcl}
     \sigma_\alpha  & = & a_\alpha + b_\alpha \, (\delta - c_\alpha) + d_\alpha/(\delta - e_\alpha)    \\
      \sigma_\delta & = & a_\delta + b_\delta \, (\delta - c_\delta) + d_\delta \, (\delta - e_\delta)^2, 
  \end{array}
\eeq{e:e5}
   where 
$a_\alpha = 0.060$~mas, $b_\alpha = 0.000011$~mas, $c_\alpha=10^\circ$, 
$d_\alpha=1$~mas, $e_\alpha = 110^\circ$; 
$a_\delta = 0.090$~mas, $b_\delta = 0.000350$~mas, $c_\delta=90^\circ$, 
$d_\delta=0.000023$~mas, $e_\delta = 60^\circ$.

  The distribution of normalized arc lengths with modified uncertainties that
accounts for the contribution of the residual ionosphere on source positions
is shown with pink pale circles in Figure~\ref{f:k_xs_raw}. The disagreement
with the Raleigh distribution is substantially reduced, but not eliminated.
That means there is another unaccounted sources of differences in source
positions.

\begin{figure}[h]
   \includegraphics[width=0.49\textwidth]{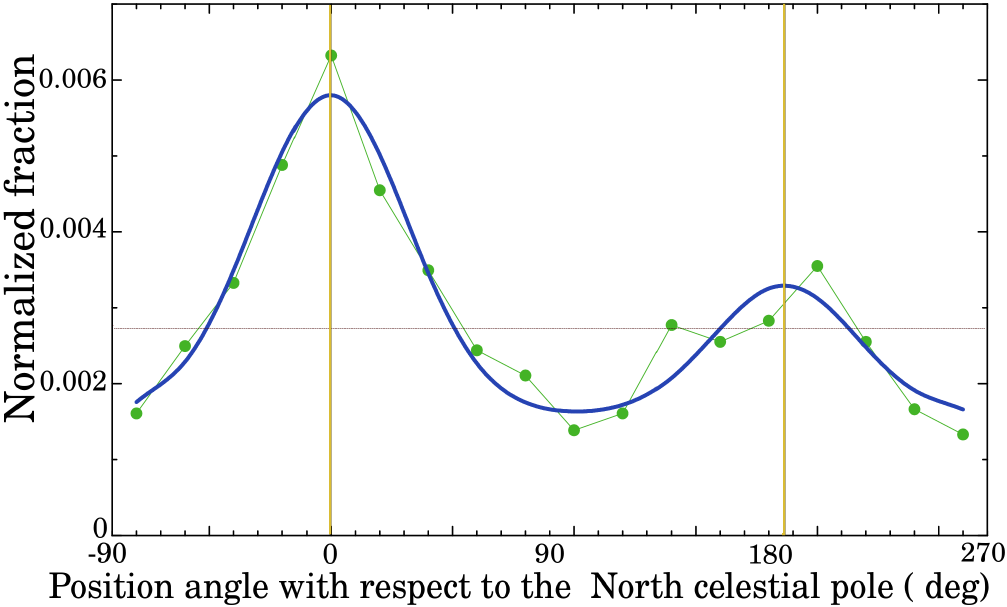}
   \caption{The normalized distribution of position angles of source
            coordinates derived from SX data counted from the North 
            celestial pole relative to the source coordinates derived
            from the K~band data solution (green circles). The solid blue 
            line shows its fit with two Gaussians and a constant. The yellow 
            vertical lines show the maxima. The thin dashed red line shows 
            the uniform distribution.
           }
   \label{f:hist_north}
\end{figure}

  Let us recollect that the position differences have two components. They
can be characterized as an arc and a position angle counted counter-clockwise
from the declination axis. Figure~\ref{f:hist_north} shows the normalized
distribution of these position angles. In the absence of systematic errors,
that histogram would have been flat (see the thin red dashed line). But the 
histogram shows two broad peaks along the declination axis, which is a 
manifestation of systematic errors. That means the position differences along 
the declination axis are more prevailing. The disparity in the peak amplitudes 
means there is a declination bias, and since the peak along $180^\circ$ is 
greater than the peak along $0^\circ$, the bias is positive.

  I fitted the empirical distribution with a simple model that consists of
a constant and a sum of two Gaussians. The maxima are at -$1^\circ$ and 
$183^\circ$ respectively. The second moments of the Gaussians turned out
to be very close: $31^\circ$ and $32^\circ$.

  All the sources observed in this campaigns were AGNs. Almost all AGNs 
exhibit morphology of a featherless core and a jet milliarcsecon scales.
I re-drew the histogram counting the position angle from jet directions. 
The jet directions were determined for all the sources used in this 
investigation by \citet{r:pla_jet}. The histogram shows two broad peaks 
at -$1^\circ$ and at $207^\circ$. Like in a case of the dependence of position
angle with respect to the North celestial pole, I fitted the histogram to 
a similar model of two Gaussians and a constant term. The primary peak along 
jet directions is 25\% higher and 10\% narrower than the secondary peak 
in the direction opposite to the jet. The contribution of the source structure 
and the residual core-shift due to a violation of the equi-partition condition 
would cause a position offset along the jet. These contributions are frequency 
dependent, and they were expected to emerge on a position angle histogram when 
position estimates from observations at different frequencies are compared. 
However, the magnitude of this effect was not known.

\begin{figure}[h]
   \includegraphics[width=0.49\textwidth]{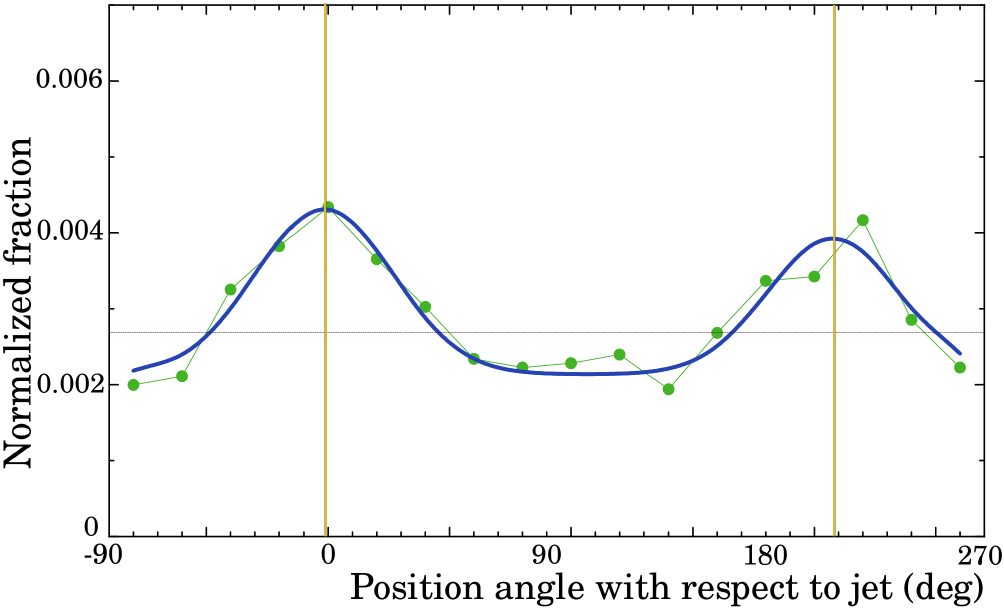}
   \caption{The normalized distribution of position angles of source
            coordinates derived from SX data counted from jet 
            directions relative to the source coordinates derived
            from the K~band data solution (green circles). The solid blue 
            line shows its fit with two Gaussians and a constant. 
            The yellow vertical lines show the maxima. The thin dashed 
            red line shows the uniform distribution.
           }
   \label{f:hist_jet}
\end{figure}

  With some extra efforts we can infer the magnitude of the extra noise along
jet directions from the histogram themselves. To perform a quantitative 
estimate and assess an uncertainty of that estimate, I performed a Monte Carlo 
simulation.

  I generated a random sequence of the simulated SX versus K~band position 
differences in a form of a sum of two terms: $R_o + R_i$. The 
first term is the observation noise. This term is the difference of two
2D Gaussian processes with the zero first moment and known second moments.
These second moments were set to uncertainties in right ascension and 
declination as well as their correlations for each source from the compared
catalogues. $R_i$ is a 2D Gaussian process with components along 
declination $A_i \cos a_i$ and along right ascension $A_i \sin a_i$, where 
$A_i = G(s_i,\sigma_{Ai})$ and $a_i = G (0,\sigma_{am})$. Here $G(a,b)$ 
denotes the Gaussian function with the first and second moments $a$ and $b$ 
respectively. This process is supposed to model the extra noise in source 
position due to the contribution of the residual ionospheric path delay. 
I fixed the second moment of the noise in angles $\sigma_{am}$ to the value 
that fit histograms Figures~\ref{f:hist_north}: $30^\circ$. I varied $s_i$ 
and $\sigma_{Ai}$ on a two-dimensional grid and sought for those $s_i$ and 
$\sigma_{Ai}$, that provide the best fit to the modeled histogram in 
Figures~\ref{f:hist_north} shown with the solid blue line. I ran 1024 trials 
in the inner loop. Then I ran the outer loop  64 times with a different seed 
of the random number generator. I got the time series of $s_i$ and 
$\sigma_{Ai}$ and computed their mean and root mean square.

  I ran a similar procedure for accounting an extra noise along jet 
directions. The position differences were represented as $R_o + R_j$.
The second term $R_j$ is the 2D Gaussian process with components along 
declination $A_j \cos (a_j + j)$ and along right ascension
$A_j \sin (a_j +j)$, where $j$ is a jet direction. I fit this stochastic
model to the modeled histogram in Figure~\ref{f:hist_jet}. I ran this
procedure two times, once with the original histogram and the second
time with the modified histogram: I changed the position of the second maximum 
from $207^\circ$ to $180^\circ$. The location of the second maximum at 
$207^\circ$ does not fit a simple model of the Gaussian noise
along jet directions. The cause of the secondary maximum shift 
is unclear. The use of the Gaussian noise along jet directions to represent
a histogram that has a secondary peak at $207^\circ$ will cause an 
underestimation of the second moment estimate. The artificial shift of the 
second maxima to $180^\circ$ eliminates that problem and provides a higher value 
of the estimate. However, since the observed histogram has the secondary maxima
at a different location, that value of the second maxima will be an 
overestimation. Therefore, these two estimates provide the upper and
lower limits. It should be noted that since the distribution of jet directions
is uniform, the source position noise driven by the residual ionospheric
path delay and the noise along jet directions are uncorrelated. That means the 
ionosphere driven noise will not affect the histogram of the jet driven noise 
and vice versus.

  Table~\ref{t:posang_est} shows results of these Monte Carlo trials.
Adding the contribution of the residual ionosphere to the contribution
of the extra noise along the jet, we get 0.130~mas. Taking the mean
uncertainty of of K and SX band position estimate differences in right 
ascension scaled by $\cos\delta$ (0.043~mas) and in declination (0.073~mas)
and comparing them with lines 4 and 6 of Table~\ref{t:k_xs}, we get an 
excess noise 0.10 and 0.14~mas over right ascension and declination.

\begin{table}
   \caption{Estimates of the bias of the ionospheric noise (upper row)
            and the noise along jet directions from fitting the 
            position angle histograms in mas.}
   \makeatletter\if@two@col \hspace{-4em} \fi\makeatother
   \begin{tabular}{lr@{\:}r@{\:}rr@{\:}r@{\:}r}
          \hline
                       & \nnntab{c}{bias} & \nnntab{c}{$\sigma$} \\
          orig north   & 0.053 & $\pm$ &  0.012 & 0.089  & $\pm$ & 0.032 \\
          orig jet     & 0.011 & $\pm$ &  0.007 & 0.093  & $\pm$ & 0.027 \\
          modified jet & 0.010 & $\pm$ &  0.007 & 0.121  & $\pm$ & 0.023 \\
         \hline
   \end{tabular}
   \label{t:posang_est}
\end{table}

  In addition, I use the histogram of normalized arc lengths between K~band
and SX band position estimates to estimate the contribution of the two origins
of the excessive noise. I represented the extra noise as 
$\lambda_i \, R_i(\delta) + \lambda_j \, R_j(j)$, where $R_i(\delta)$ is 
the ionospheric noise according to the regression expression~\ref{e:e5}, and
$R_j(j)$ is the same process as used for the fitting of the histogram with 
respect to jet directions with $s_j = 0.011$~mas and 
$\sigma_{Ai}=0.093$~mas taken from Table~\ref{t:posang_est}. I considered 
$\lambda_i$ and $\lambda_j$ as free dimensionless admittance factors. 
I evaluated $\lambda_i$ and $\lambda_j$ from a broad range [0.2, 2.0] with 
a step of 0.01 to provide the best fit to the histogram of normalized arc 
lengths. In order to provide a rough estimate of the uncertainty of that 
fit, I varied the number of bins in the histogram in a range of 15 to 25 
and computed the mean and the rms of the estimates. The estimates are 
$\lambda_i = 1.10 \pm 0.07$ and $\lambda_i = 0.91 \pm 0.11$.

\begin{figure}[h]
   \includegraphics[width=0.49\textwidth]{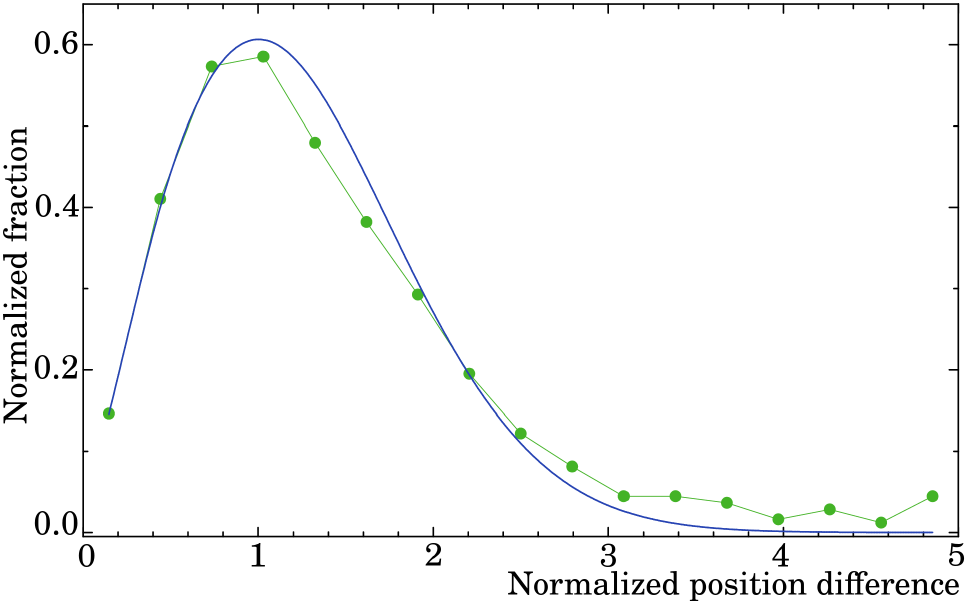}
   \caption{The distribution of normalized arc lengths between K band 
            and SX source position estimates with the contribution from 
            the residual ionosphere and an extra noise along jet 
            directions applied with the admittance factors equal to one. 
            The thick blue line shows the Rayleigh distribution 
            with $\sigma=1$ as a reference.
           }
   \label{f:k_xs_fitted}
\end{figure}

  Although both approaches, fitting the histogram of position angles and 
fitting the histogram of normalized arc lengths, uses the same source 
position catalogues, these results are independent. We can consider 
an arc length and a position angle as two components of the position 
differences. Therefore, the agreement within one standard deviation of 
these two approaches is very encouraging. Figure~\ref{f:k_xs_fitted} 
shows the histogram of normalized arc lengths with the extra noise 
due to the residual ionospheric contribution and the extra noise 
along jet directions applied with the admittance factors equal to unity. 
Compare it with Figure~\ref{f:k_xs_raw}.

  Finally, to investigate the impact of scattering in the interstellar medium,
I reran the comparison by including and excluding observations with galactic 
latitude less than $5^\circ$. I included sources with position differences
up to 1~mas in this comparison. I found that the agreement in the area 
within $5^\circ$ of the Galactic plane is worse by 34\% in right ascension 
and 37\% in declination with respect to the area above the plane. 
I attribute these differences to an increase of source position errors 
from 2.2/8.4~GHz data with respect to positions from 23.6~GHz that are
mostly not affected.

\subsection{Comparison of Q~band positions against SX positions}

  I compared source positions derived from Q-band against those derived
from K~band. The comparison showed a bias -0.14~mas over declination.
The rms of position differences of 284 compact sources is 0.25~mas over
right ascension scaled by $\cos\delta$ and 0.42~mas over declination, 
which is about 60\% greater than median Q-band position errors after 
adding an extra noise in quadrature to make the errors over right 
ascension and declination Gaussian: 0.17 and 0.34~mas respectively.

\section{Discussion}

  Table~\ref{t:difs} summarizes the differences in source position 
estimates. The first five rows show results of comparison of source 
position determination using dual- and quad-band VLBI observations. 
Comparison of positions from quad-band data versus dual-band data 
shows that the extra variance in positions unaccounted in reported 
errors is at a level of 0.07~mas. This noise does not depend on 
a choice of frequency bands: the use of 2.2/8.4, 4.1/7.4 or quad-band 
data do not introduce discernible systematic errors. Comparison of 
positions from CX data at 4.1/7.4~GHz shows a noticeably greater scatter,
but no systematic pattern. It should be noted that CX VLBA dataset 
analyzed here originates from different programs that have been 
scheduled in a different way than all other programs in this study. 
Bright sources were observed as calibrators, and the schedules were 
not optimized to reach the highest position accuracy of calibrators. 
Nevertheless, the extra noise for the comparison with source positions 
from CX data is on par with other catalogue pairs.

\begin{table*}
   \caption{Statistics of the arc length differences. 
            Column 1: the number of sources used in comparison; 
            column 2: declination bias; 
            column 3: rms over right ascension scaled by $\cos\delta$;  
            column 4: rms over declination; 
            column 5: extra noise in right ascension; 
            column 6: extra noise in right declination; 
            column 7: extra variance along jet directions. Units are mas.
           }
   \makeatletter\if@two@col \hspace{-4em} \fi\makeatother
   \small
   \begin{tabular}{l r rr rr rr}
       \hline
                                              &  (1) &  (2)   & (3)   & (4)   & (5)   & (6)   & (7)   \\
       {\sc vlba} SX/{\sc ivs VGOS} quad      &  204 & -0.056 & 0.106 & 0.155 & 0.098 & 0.090 & ---   \\  
       {\sc ivs r1r4} SX /{\sc ivs VGOS} quad &  197 &  0.009 & 0.086 & 0.129 & 0.068 & 0.075 & ---   \\ 
       {\sc vlba} SX/{\sc ivs r1r4} SX        &  665 &  0.061 & 0.125 & 0.153 & 0.060 & 0.091 & 0.111 \\ 
       {\sc vlba+} SX/{\sc ivs r1r4} SX       &  653 &  0.020 & 0.089 & 0.108 & 0.061 & 0.062 & 0.076 \\ \vex  
       {\sc vlba} CX/{\sc vlba} SX            &  734 &  0.030 & 0.266 & 0.387 & 0.112 & 0.092 & ---   \\ 
       {\sc vlba} SX/{\sc vlba} K             &  848 & -0.042 & 0.126 & 0.182 & 0.082 & 0.107 & 0.093 \\
       {\sc vlba} Q/{\sc vlba} K              &  284 & -0.160 & 0.245 & 0.170 & 0.300 & 0.426 & ---   \\
       {\sc ivs r1r4} SX/{\sc vlba} K         &  526 &  0.106 & 0.119 & 0.194 & 0.035 & 0.070 & 0.080 \\
       {\sc vlba} X/{\sc vlba} SX             & 4624 &  0.058 & 0.265 & 0.383 & 0.210 & 0.250 & 0.058 \\
       {\sc vlba} X/{\sc vlba} K              &  848 & -0.040 & 0.191 & 0.256 & 0.083 & 0.128 & 0.088 \\
       \hline
   \end{tabular}
   \label{t:difs}
\end{table*}

  Some comparisons revealed declination biases at a level comparable with
the accuracy of source position catalogues. Since these biases emerged also 
in the comparisons of source positions derived from data at the same bands, 
the origin of these biases is not related to the frequency selection. It 
should be noted that there is no biases from source positions derived at 
global networks used in R1R4 and VGOS campaigns. There is a systematic bias 
between positions derived from the VLBA network with respect to positions 
derived from the global network. Mismodeling atmospheric path delay is 
the factor that may play a certain role. These errors are elevation dependent. 
A given source is observed in a wider range of elevations at a global 
network. This mitigates systematic errors. To check this conjecture, I made 
an extended solution using SX data from VLBA, including all the stations 
outside the VLBA network that participated in experiments. I designated that 
solution as VLBA+. The declination bias dropped from 0.06~mas to 0.02~mas, 
and the additional noise to make the residual distribution close to the 
Gaussian reduced from 0.09 to 0.06~mas. 

  An angle of 0.07~mas, or 0.3~nrad, corresponds to 2~mm on the Earth. 
Geodetic VLBI experiments usually observe $\sim\! 100$~sources during 
24~hr periods. If source position errors are totally uncorrelated, their 
impact on geodetic results will be at a level of 0.2~mm. Considering that 
the current accuracy of determination of the horizontal station position 
components is at a level of 1--3~mm from a 24~hr experiment, this is not 
a concern. However, the declination bias of 0.06~mas can potentially cause
the latitude bias up to 2~mm. During  geodetic analysis we either estimate 
source positions from geodetic observations or fix them to positions 
determined from data analysis at other networks. We should keep in mind 
that fixing source positions potentially may introduce systematic errors 
up to 2~mm.

\begin{figure}[h]
   \includegraphics[width=0.49\textwidth]{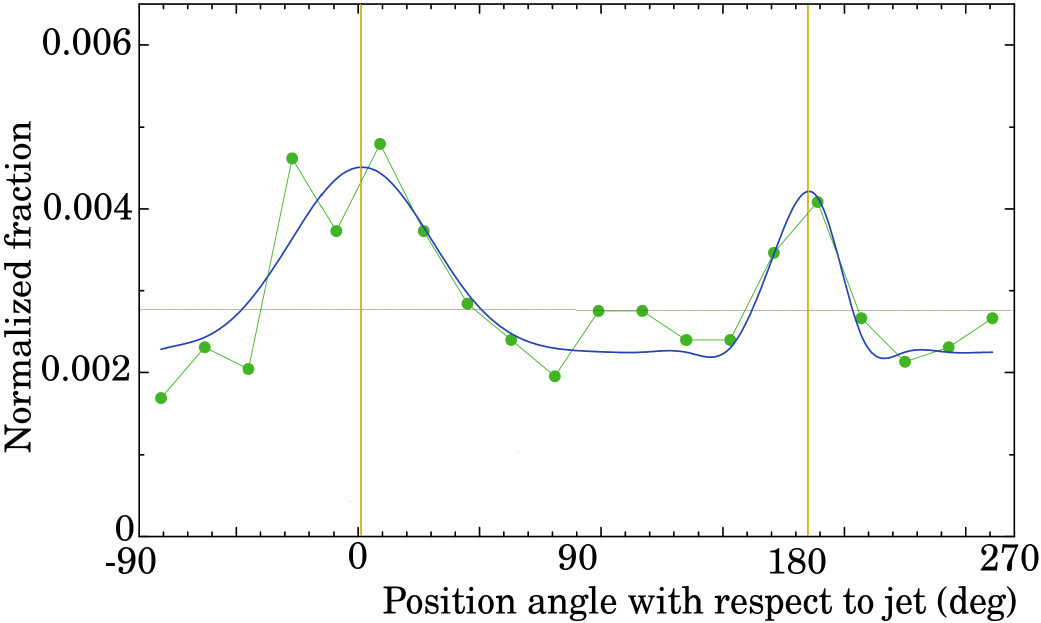}
   \caption{The normalized distribution of position angles of source
            coordinates derived from VLBA SX data counted from jet 
            directions relative to source coordinates derived from 
            IVS R1R4 SX data (green circles). The solid blue line shows 
            its fit with two Gaussians and a constant. The yellow vertical 
            lines show the maxima. The thin dashed red line shows the 
            uniform distribution.
           }
   \label{f:hist_r1r4_xs_jet}
\end{figure}

\begin{figure}[h]
   \includegraphics[width=0.49\textwidth]{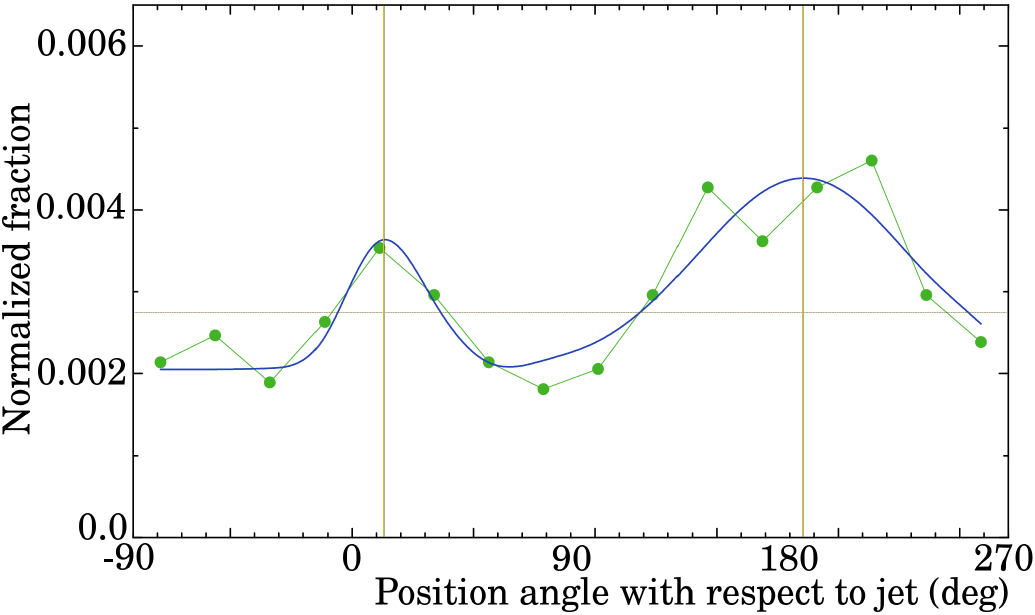}
   \caption{The normalized distribution of position angles of source
            coordinates derived from IVS R1R4 SX~band data counted from 
            jet directions relative to the source coordinate from VLBA 
            K~band data (green circles). The solid blue line shows its 
            fit with two Gaussians and a constant. The yellow vertical 
            lines show the maxima. The thin dashed red lines shows the 
            uniform distribution.
           }
   \label{f:hist_r1r4_k_jet}
\end{figure}

  The histograms of the position angles of source position differences with
respect to jet directions showed peaks in the direction along the jet.
We can identify these peaks when compare source positions from IVS R1R4 data
as well --- see Figures~\ref{f:hist_r1r4_xs_jet}--\ref{f:hist_r1r4_k_jet}.
The presence of this pattern shows unambiguously that source structure and 
core shift affect source positions. Which of these two effects is dominant?
The core-shift manifests as a displacement of the core with respect to the 
black hole. It does not dependent on the network. If it were the core-shift, 
we would expect to see a pattern in the position angle histogram for solutions 
at different frequencies, 8 and 23~GHz, but not in positions derived from 
observations at the same frequencies. Because of the core-shift variability, 
we still may see the residuals of the effect in source position differences 
if campaign are observed in different time epochs, but the residuals are 
expected to be small. In contrast, source structure is network dependent. 
Considering that the magnitude of the noise along jet directions between SX 
and K band solutions at the VLBA network is comparable with the magnitude 
between SX solutions at the VLBA and IVS network, an explanation of the peaks 
at the histograms by the contribution of the unaccounted source structure 
is much more plausible. The magnitude of the contribution is consistent
with the a~priori estimate based on analysis of 15~GHz VLBA observations
presented in \citet{r:gaia3}: 0.06~mas.

  It should be noted that a dataset should be rather large in order to 
identify peaks in a position angle histogram. A histogram computed from
665 points with 20 bins has on average only 33 points per bin. 
Histograms computed using source position differences derived from VGOS 
datasets are too noisy to see peaks there because on average, they have 
only 10 points per bin.

  At which frequency the impact of source structure is stronger? We can 
examine three hypotheses: the impact is stronger at 8~GHz, the impact is 
about the same, and the impact is stronger at 23~GHz. Since the 
contribution of the extra noise along jet directions based on the comparison 
of the differences in source positions derived from dual-band observations 
at 2.2/8.4 versus single-band observations at 23~GHz is comparable with the 
contribution of the extra noise from the differences in source positions 
derived from observations at the same frequency, we can rule out the third 
hypothesis. Analysis of the position differences between 43 and 23~GHz 
potentially could have helped, but unfortunately, the position differences are 
too noisy to detect the extra noise along jet directions. We have to conclude 
that the presented data analysis does not to allow us to discriminate two 
other hypotheses. 

  Comparison of positions of source estimates derived from single-band VLBA 
observations at 8~GHz against position estimates derived from dual-band 
observations revealed that positions have an extra noise at a level of 
0.2--0.3~mas. The total differences are at a level of 0.3--0.4~mas. It should 
be noted that these estimates are average over several solar cycles. It is 
instructive to compare these estimates with the estimates of processing 5~GHz 
VLBA data collected during the solar minimum that had the median extra noise 
0.5~mas \citep{r:astro_vips} and processing 8~GHz data with the Australian 
Long Baseline Array (LBA) --- that had the extra noise of 3.2~mas 
\citep{r:lcs2}. The one order of magnitude difference between VLBA and LBA 
results is quite large. Large errors from LBA single-band observations were 
attributed to large errors in the ionospheric model derived from GNSS, but 
this explanation may need be revised considering results of the present study.

  The error budget in single-band observations consists of the contribution
of the frequency-independent error floor, the source structure contribution,
and the ionospheric contribution. The prior analysis showed that each 
constituent is in a range of 0.05--0.07~mas per source position component, 
except the contribution of the ionosphere to the declination noise, which is 
greater at declination below $0^\circ$. Observed differences in source 
positions are fully consistent with these three constituents on 
a {\it quantitative} level, which is exciting. This analysis does not 
address the nature of the frequency-independent error floor. Investigation 
of this error floor is beyond the scope of this study and will be addressed 
in a separate publication. 

  I should stop short of declaring that these three phenomena fully explain 
the differences. We can establish causality if we can model a phenomenon,
include it in the data reduction algorithm, and achieve a reduction of 
residuals. The reduction of residuals at a level prescribed by the model
establishes a necessary and sufficient criteria for the phenomena in
consideration to explain the measurement results. This sets the standard of 
problem solving. Establishing a stochastic model described by two first 
moments of their distribution density allows us to establish only 
a sufficient condition. We can say that these three contributions are 
sufficient for explaining the results at the current level of accuracy, 
and no further phenomena are necessary to include.

\begin{figure}[h]
   \includegraphics[width=0.49\textwidth]{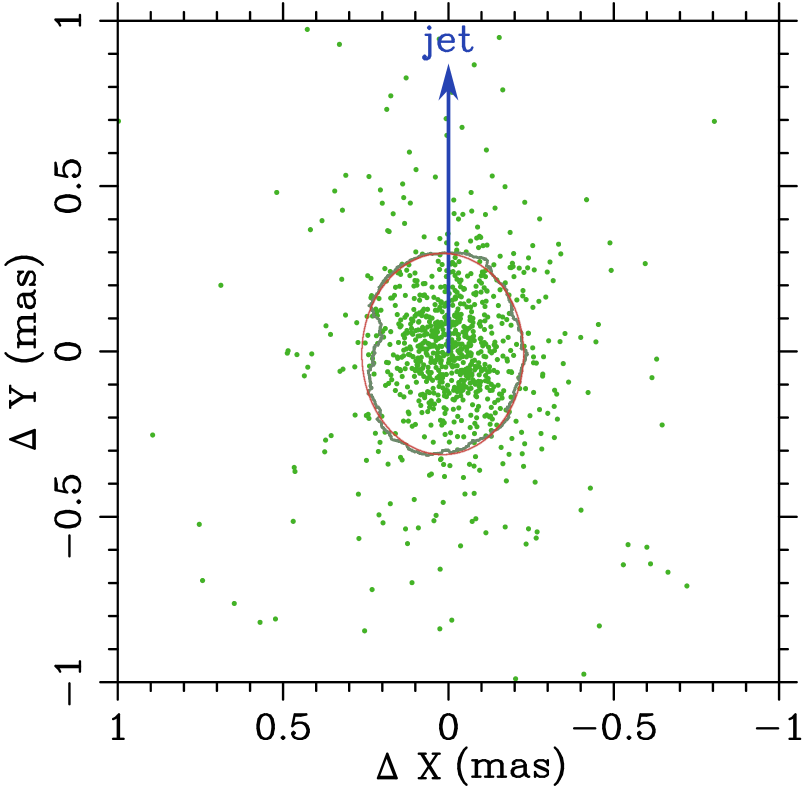}
   \caption{The differences in positions from SX~band VLBA data with 
            respect to positions from K~band data. The vertical axis 
            is aligned with jet directions. The ragged gray line shows 
            the average position offset in a sector of $90^\circ$. 
            The thin smooth red line shows its approximation with 
            an ellipse.
           }
   \label{f:k_xs_rose}
\end{figure}

  Although the presence of the noise aligned with jet directions is 
firmly established from the analysis of the position differences, we should 
keep in mind that this phenomenon does not dominate the residuals. 
Figure~\ref{f:k_xs_rose} shows the diagram of two-dimensional differences
in position estimates. The thick gray line shows the average position offset
with a sector of $\pm 45^\circ$. The thin smooth red line approximates the 
ragged gray line with an ellipse. The thin smooth line be circular in the 
absence of the source structure contribution. It is just its flattening 
(0.2) and an offset that allows us to make an inference about the presence 
of the source structure contribution as a Gaussian stochastic process 
and to assess its first and second moments.

  A true stochastic model may be rather complicated. The most 
simple form of such a model is $\sigma^2 = R \, \sigma^2_{\rm orig} + n^2$.
That form was used in the past, f.e. for deriving ICRF catalogues 
\citep{r:icrf3}. Even in that simplistic formulation, it is problematic to
separate reliably a multiplicative factor $R$ from an additive noise with 
a variance of $n^2$ when fitting a histogram. The distribution of the 
normalized arc lengths of position estimate differences from data at 
23.6 and 2.2/8.4~GHz can fit reasonably well to the Rayleigh 
distribution either by adding in quadrature the variance or by multiplying 
residuals by a factor of 1.82. The use of an additive stochastic model 
implies that noise contributions are independent. Indeed, it 
is reasonable to consider that the thermal noise in receivers, the noise 
due to mismodeling path delay in the ionosphere, in the neutral atmosphere, 
and the contribution of source structure as independent, since we do not 
have evidence that these processes could be correlated. An explanation of 
appearance of multiplicative factors is problematic. The multiplicative 
factors can potentially explain an intrinsic measurement noise of a given 
source position catalogue. But in that case, we expect the same intrinsic 
noise to affect results in decimation tests. The decimation tests for all 
position catalogues, except the one derived from VGOS observations, 
provided the estimates of $R$ factors close to unity, which contradicts 
to an assumption of the presence of a significant multiplicative factor 
in position uncertainties.

\begin{figure*}
   \includegraphics[width=0.33\textwidth]{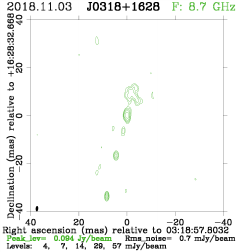}
   \includegraphics[width=0.32\textwidth]{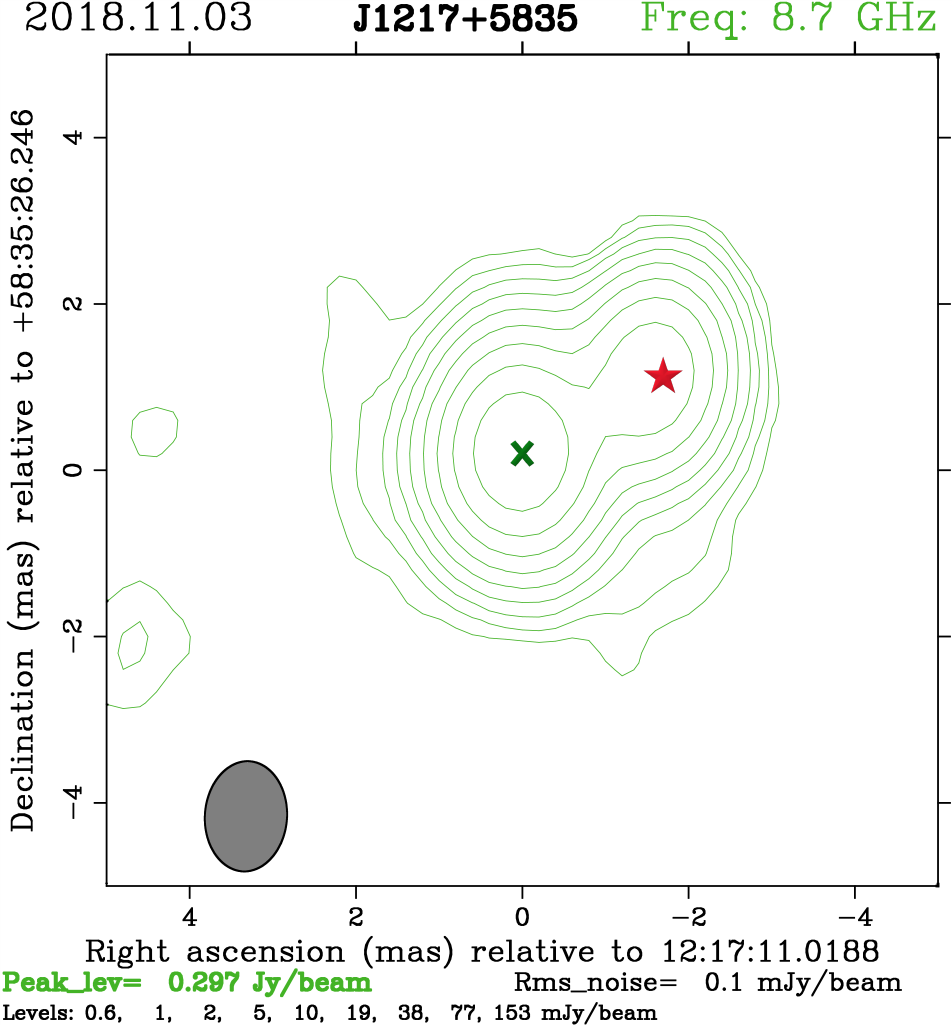}
   \includegraphics[width=0.36\textwidth]{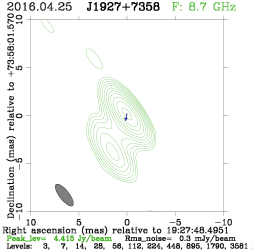}
   \caption{Image of the sources with large positions differences
            between coordinates derived from  8.4/2.3~GHz 
            data with respect to coordinates derived from 23.6~GHz data.
            {\it Left: } J0318+1628 also known as CTA21 (position 
            difference -5.10~mas in right ascension);
            {\it Center: } J1217+5835. Position estimate at 2.2/8.4~GHz
                         is shown with a green cross, position estimate
                         at 23.6~GHz is shown with a red star;
            {\it Right: } J1927+7358. The offset of position from 
                          2.2/8.4~GHz data relative of its position
                          estimate at 23.6~GHz is shown with 
                          a blue arrow.
           The images are taken from the Astrogeo VLBI FITS image database
           available at \web{http://astrogeo.smce.nasa.gov/vlbi\_images}
          }
   \label{f:source_maps}
\end{figure*}

 Figures~\ref{f:xsxc_ra},\ref{f:xsxc_dec},\ref{f:vgos_r1r4_ra},\ref{f:vgos_r1r4_dec},\ref{f:xs_r1r4_ra},\ref{f:xs_r1r4_dec}
of source position differences had a small range to focus on the bulk of 
the sources. There was a number of sources that were left beyond the bounding
box of the plots. The total number of outliers between dual-band, quad-band,
or single-band observations exceeding $3\sigma$ was at a level of 6\% and 
exceeding $5\sigma$ was at a level of 2\%. A close examination of images 
usually allows us to reveal easily the cause in most of the cases. Three 
most common cases are shown in Figure~\ref{f:source_maps}. J0318+1628 with 
the largest offset over right ascension (\mbox{-5.10~mas}) with respect to 
the position estimate at 23.6~GHz has three compact components on its jet. 
An example of sources with significant position differences when observed 
at different frequencies was reported for the first time in \citet{r:vgaps}.
Later, more sources like those have been found 
\citep[for instance,][]{r:obrs2,r:min22}. J1217+5835 shown in the 
center has two brightness peaks on its image at 8.7~GHz, A and B, shown with 
a green cross and a red star respectively. The position estimate from 
2.2/8.4~GHz corresponds to component A that is the brightest at 8.4~GHz. 
The 23.6~GHz image \citep[see Figure~3 in][]{r:witt22} reveals that component 
B is the brightest and the most compact at 23.6~GHz. The image of J1927+7358
in the right exhibits a jet. The position offset 0.6~mas along the jet is shown
with a small blue arrow. This third case is less common and can be considered 
as a tail of the distribution of position offsets along the jet caused 
by unaccounted source structure.

\section{Summary, conclusions, outcomes, and perspectives}

  I have processed seven extended datasets that include single-band, dual-band, 
and quad-band observations at different frequencies and different networks, 
derived source positions using the absolute astrometry approach, examined 
their differences, and fitted a stochastic model that reconcile the normalized 
residuals to either the Gaussian or the Rayleigh distributions. The most 
important findings are the following:

\begin{enumerate}

  \item The level of the agreement between source position estimates derived 
        from quad-band, dual-band, and single-band observations is below
        a 1~nrad level, or 0.2~mas. The only systematic error found was 
        a declination bias at a level of 0.02--0.10~mas. Since such a bias 
        was found also in analysis of data collected at different networks, 
        this bias is not considered to be related to a selection of observing 
        frequencies.

  \item Comparisons revealed that the number of outliers is at a level of 
        2 to 6\% that corresponds to a normalized position difference 
        $5\sigma$ or $3\sigma$. The upper level of that range over 
        normalized position differences corresponds to the sources that 
        are definitely peculiar, the lower level corresponds to the sources 
        that might be peculiar. Most of the peculiar sources have more 
        than one bright component in their images.
  
  \item The source position catalogues derived from dual-band, quad-band, 
        and single band observations at 23.6~GHz has three constituents of 
        the extra noise with approximately equal second moment 0.05--0.07~mas 
        per position component. These components are 1)~the intrinsic common 
        noise, probably caused by mismodeling path delay in the neutral 
        atmosphere; 2)~the noise caused by mismodeling the ionospheric 
        contribution when processing single-band data, and 3)~the noise 
        predominately along jet directions caused by mismodeling of 
        source structure and possibly, by the core-shift. The latter noise 
        manifests itself in a histogram of the position angle of the 
        differences in source coordinate estimates with respect to jet 
        directions in a form of two broad peaks at $0^\circ$ and $180^\circ$.
        The astrometric analysis does not allow us to determine at which band 
        the contribution of source structure is the greatest, but it allows 
        us to rule out the hypothesis that source structure contribution at 
        23.6~GHz is greater than at 8.4~GHz.

  \item The error budget of the differences between dual-band 2.2/8.4 and 
        23.6~GHz observations was established on a quantitative level. 
        The source position catalogues have a common intrinsic noise level 
        with the second moment of around 0.05~mas per component, a Gaussian 
        noise along jet directions with the second moment of 0.09~mas that 
        definitely affects position estimates at low frequencies and may affect 
        position estimates at high frequencies, and the residual ionospheric 
        noise that affects only position estimates from high frequency data. 
        It contributes at a level of 0.07~mas to declination above $0^\circ$, 
        grows with a decrease of declination, and reaches 0.3~mas at 
        declination $-45^\circ$. It is sufficient to consider these three 
        phenomena in order to close the error budget.

  \item Position differences derived from 2.2/8.4 and 23.6~GHz in the area of 
        $5^\circ$ the Galactic plane are 1/3 higher than otherwise. This 
        difference is attributed to an improvement of source position
        accuracy at 23~GHz because observations at high frequencies are less 
        affected by scattering in the interstellar medium.
\end{enumerate}

  A frequency-dependent noise in a dual-band or quad-bands setup or in 
a single-band setup at 22~GHz or above affects source position estimates 
at a level of not exceeding 0.07~mas per component. The random noise in 
source positions of that magnitude affects station position estimates at 
a level of 0.2~mm, provided all source position estimates are uncorrelated. 
Source position catalogues may have declination biases up to 0.1~mas, but 
these biases are present in estimate of source positions not only derived 
from observations at different frequencies, but from observations at 
different networks as well, especially at networks of a small size. 
Expanding the network to a global scale reduces these biases to 
a negligible level. Possible biases in source positions derived from 
observations at regional networks should be checked when fixed source 
positions are used in data analysis.

  Analysis of source positions differences suggests that observations at 
23~GHz has a potential of reducing the contribution of source structure 
that manifests itself as the Gaussian noise along jet directions with the 
second moment 0.07--0.09~mas. However, the magnitude of that reduction 
is not yet known. At the same time, the residual ionospheric noise affects
only source position estimates derived from 23~GHz data, and this noise is 
comparable with the extra noise due to source structure.
  
  Presented analysis provides enough evidence to conclude that in general, 
absolute radio astrometry at 23~GHz currently cannot outperform astrometry 
at 2.2/8.4, 4.1/7.4, or 3.0/5.2/6.4/10.2~GHz. K~band astrometry has 
a potential to outperform CX or SX astrometry only in the sky areas 
with high scattering, i.e. its role is rather marginal. This consideration 
should be taken into account for planning future observing programs. 

  Presented analysis provides firm evidence that a development of 
frequency-dependent celestial reference frame is currently not warranted.
High-frequency observations in the areas with a high density of the 
interstellar medium may significantly improve position accuracy. Observations 
in these areas at 22~GHz and above are highly desirable. This is the main 
niche of high-frequency radioastrometry.

  Since frequency-dependent source position errors, 0.05--0.07~mas per 
component, are comparable with common errors that affect all catalogues 
and with network-dependent errors, the use of dual-band, quad-band, and 
signal-band data at 22~GHz and above in a single least square solution 
will cause an additional error not exceeding 0.07~mas. This conclusion
should also be taken into account for designing future  astrometric programs.

   Considering future perspectives, ionospheric errors can be mitigated either
by observing simultaneously at 8/23 or 8/32~GHz or by solving for biases 
in the global ionospheric maps derived from GNSS observations following the 
technique presented in \citet{r:sba} with the use of data from collocated
GNSS stations. If the residual ionospheric contribution will be entirely 
eliminated, the position accuracy can potentially be improved from a level 
of 0.07~mas to a 0.05~mas level. 

\begin{acknowledgments}
   This work was done using only publicly available datasets 1)~collected 
with the VLBA network of the NRAO and available at 
\web{https://data.nrao.edu/portal/} and 2)~collected with the IVS network and 
available at the NASA Crustal Dynamics Data Informational System (CDDIS) 
\web{https://cddis.nasa.gov/archive/vlbi/}. The NRAO is a facility 
of the National Science Foundation operated under cooperative agreement by
Associated Universities, Inc. The author acknowledges use of the VLBA under 
the USNO's time allocation for some datasets. This work made use of the 
Swinburne University of Technology software correlator, developed as part of 
the Australian Major National Research Facilities Programme and operated 
under license. This work was supported by NASA's Space Geodesy Project.                
\end{acknowledgments}

   It is my pleasure to thank Yuri Y.~Kovalev for thoughtful discussions.
The author would like to thank Chris Jacobs for a constructive criticism of
early results and an encouragement to perform a quantitative study of the error 
budget of high frequency absolute radio astrometry.

\facility{VLBA,IVS}
\software{PIMA,pSolve}

\bibliographystyle{aasjournal}
\bibliography{radf}

\begin{thebibliography}{}
\expandafter\ifx\csname natexlab\endcsname\relax\def\natexlab#1{#1}\fi
\providecommand{\url}[1]{\href{#1}{#1}}
\providecommand{\dodoi}[1]{doi:~\href{http://doi.org/#1}{\nolinkurl{#1}}}
\providecommand{\doeprint}[1]{\href{http://ascl.net/#1}{\nolinkurl{http://ascl.net/#1}}}
\providecommand{\doarXiv}[1]{\href{https://arxiv.org/abs/#1}{\nolinkurl{https://arxiv.org/abs/#1}}}

\bibitem[{{Abell{\'a}n} {et~al.}(2018){Abell{\'a}n}, {Mart{\'\i}-Vidal},
  {Marcaide}, \& {Guirado}}]{r:abe18}
{Abell{\'a}n}, F.~J., {Mart{\'\i}-Vidal}, I., {Marcaide}, J.~M., \& {Guirado},
  J.~C. 2018, \aap, 614, A74, \dodoi{10.1051/0004-6361/201731869}

\bibitem[{{Campbell} {et~al.}(1988){Campbell}, {Schuh}, \&
  {Zeppenfeld}}]{r:zep88}
{Campbell}, J., {Schuh}, H., \& {Zeppenfeld}, G. 1988, in The Impact of VLBI on
  Astrophysics and Geophysics, ed. M.~J. {Reid} \& J.~M. {Moran}, Vol. 129, 427

\bibitem[{{Charlot}(2002)}]{r:cha02}
{Charlot}, P. 2002, in International VLBI Service for Geodesy and Astrometry:
  General Meeting Proceedings, ed. N.~R. {Vandenberg} \& K.~D. {Baver}, 233

\bibitem[{{Charlot} {et~al.}(2010){Charlot}, {Boboltz}, {Fey}, {Fomalont},
  {Geldzahler}, {Gordon}, {Jacobs}, {Lanyi}, {Ma}, {Naudet}, {Romney},
  {Sovers}, \& {Zhang}}]{r:kq2}
{Charlot}, P., {Boboltz}, D.~A., {Fey}, A.~L., {et~al.} 2010, \aj, 139, 1713,
  \dodoi{10.1088/0004-6256/139/5/1713}

\bibitem[{Charlot {et~al.}(2020)Charlot, Jacobs, Gordon, Lambert, de~Witt,
  B\"{o}hm, Fey, Heinkelmann, Skurikhina, Titov, Arias, Bolotin, Bourda, Ma,
  Malkin, Nothnagel, Mayer, MacMillan, Nilsson, \& Gaume}]{r:icrf3}
Charlot, P., Jacobs, C.~S., Gordon, D., {et~al.} 2020, \aap, 644, A159,
  \dodoi{10.1051/0004-6361/202038368}

\bibitem[{{de Witt} {et~al.}(2023{\natexlab{a}}){de Witt}, {Jacobs}, {Gordon},
  {Bietenholz}, {Nickola}, \& {Bertarini}}]{r:witt23}
{de Witt}, A., {Jacobs}, C.~S., {Gordon}, D., {et~al.} 2023{\natexlab{a}}, \aj,
  165, 139, \dodoi{10.3847/1538-3881/aca012}

\bibitem[{{de Witt} {et~al.}(2023{\natexlab{b}}){de Witt}, {Jacobs}, {Gordon},
  {Hunt}, \& {Johnson}}]{r:witt22}
{de Witt}, A., {Jacobs}, C.~S., {Gordon}, D., {Hunt}, L., \& {Johnson}, M.
  2023{\natexlab{b}}, in International VLBI Service for Geodesy and Astrometry
  2022 General Meeting Proceedings, ed. K.~L. {Armstrong}, D.~{Behrend}, \&
  K.~D. {Baver}, 293--297

\bibitem[{{Deller} {et~al.}(2007){Deller}, {Tingay}, {Bailes}, \&
  {West}}]{r:difx1}
{Deller}, A.~T., {Tingay}, S.~J., {Bailes}, M., \& {West}, C. 2007, \pasp, 119,
  318, \dodoi{10.1086/513572}

\bibitem[{{Deller} {et~al.}(2011){Deller}, {Brisken}, {Phillips}, {Morgan},
  {Alef}, {Cappallo}, {Middelberg}, {Romney}, {Rottmann}, {Tingay}, \&
  {Wayth}}]{r:difx2}
{Deller}, A.~T., {Brisken}, W.~F., {Phillips}, C.~J., {et~al.} 2011, \pasp,
  123, 275, \dodoi{10.1086/658907}

\bibitem[{{Gordon} {et~al.}(2016){Gordon}, {Jacobs}, {Beasley}, {Peck},
  {Gaume}, {Charlot}, {Fey}, {Ma}, {Titov}, \& {Boboltz}}]{r:vcs-ii}
{Gordon}, D., {Jacobs}, C., {Beasley}, A., {et~al.} 2016, \aj, 151, 154,
  \dodoi{10.3847/0004-6256/151/6/154}

\bibitem[{Gray \& Allan(1974)}]{r:tri_corner_hat}
Gray, J., \& Allan, D. 1974, in 28th Annual Symposium on Frequency Control,
  243--246, \dodoi{10.1109/FREQ.1974.200027}

\bibitem[{{Hawarey} {et~al.}(2005){Hawarey}, {Hobiger}, \& {Schuh}}]{r:iono2nd}
{Hawarey}, M., {Hobiger}, T., \& {Schuh}, H. 2005, \grl, 32, L11304,
  \dodoi{10.1029/2005GL022729}

\bibitem[{{Kim} {et~al.}(2020){Kim}, {Krichbaum}, {Broderick}, {Wielgus},
  {Blackburn}, {G{\'o}mez}, {Johnson}, {Bouman}, {Chael}, {Akiyama}, {Jorstad},
  {Marscher}, {Issaoun}, {Janssen}, {Chan}, {Savolainen}, {Pesce}, {{\"O}zel},
  {Alberdi}, {Alef}, {Asada}, {Azulay}, {Baczko}, {Ball}, {Balokovi{\'c}},
  {Barrett}, {Bintley}, {Boland}, {Bower}, {Bremer}, {Brinkerink},
  {Brissenden}, {Britzen}, {Broguiere}, {Bronzwaer}, {Byun}, {Carlstrom},
  {Chatterjee}, {Chatterjee}, {Chen}, {Chen}, {Cho}, {Christian}, {Conway},
  {Cordes}, {Crew}, {Cui}, {Davelaar}, {De Laurentis}, {Deane}, {Dempsey},
  {Desvignes}, {Dexter}, {Doeleman}, {Eatough}, {Falcke}, {Fish}, {Fomalont},
  {Fraga-Encinas}, {Friberg}, {Fromm}, {Galison}, {Gammie}, {Garc{\'\i}a},
  {Gentaz}, {Georgiev}, {Goddi}, {Gold}, {G{\'o}mez-Ruiz}, {Gu}, {Gurwell},
  {Hada}, {Hecht}, {Hesper}, {Ho}, {Ho}, {Honma}, {Huang}, {Huang}, {Hughes},
  {Ikeda}, {Inoue}, {James}, {Jannuzi}, {Jeter}, {Jiang}, {Jimenez-Rosales},
  {Jung}, {Karami}, {Karuppusamy}, {Kawashima}, {Keating}, {Kettenis}, {Kim},
  {Kim}, {Kino}, {Koay}, {Koch}, {Koyama}, {Kramer}, {Kramer}, {Kuo}, {Lauer},
  {Lee}, {Li}, {Li}, {Lindqvist}, {Lico}, {Liu}, {Liuzzo}, {Lo}, {Lobanov},
  {Loinard}, {Lonsdale}, {Lu}, {MacDonald}, {Mao}, {Markoff}, {Marrone},
  {Mart{\'\i}-Vidal}, {Matsushita}, {Matthews}, {Medeiros}, {Menten}, {Mizuno},
  {Mizuno}, {Moran}, {Moriyama}, {Moscibrodzka}, {Musoke}, {M{\"u}ller},
  {Nagai}, {Nagar}, {Nakamura}, {Narayan}, {Narayanan}, {Natarajan}, {Neri},
  {Ni}, {Noutsos}, {Okino}, {Olivares}, {Ortiz-Le{\'o}n}, {Oyama}, {Palumbo},
  {Park}, {Patel}, {Pen}, {Pi{\'e}tu}, {Plambeck}, {PopStefanija}, {Porth},
  {Prather}, {Preciado-L{\'o}pez}, {Psaltis}, {Pu}, {Ramakrishnan}, {Rao},
  {Rawlings}, {Raymond}, {Rezzolla}, {Ripperda}, {Roelofs}, {Rogers}, {Ros},
  {Rose}, {Roshanineshat}, {Rottmann}, {Roy}, {Ruszczyk}, {Ryan}, {Rygl},
  {S{\'a}nchez}, {S{\'a}nchez-Arguelles}, {Sasada}, {Schloerb}, {Schuster},
  {Shao}, {Shen}, {Small}, {Sohn}, {SooHoo}, {Tazaki}, {Tiede}, {Tilanus},
  {Titus}, {Toma}, {Torne}, {Trent}, {Traianou}, {Trippe}, {Tsuda}, {van
  Bemmel}, {van Langevelde}, {van Rossum}, {Wagner}, {Wardle}, {Ward-Thompson},
  {Weintroub}, {Wex}, {Wharton}, {Wong}, {Wu}, {Yoon}, {Young}, {Young},
  {Younsi}, {Yuan}, {Yuan}, {Zensus}, {Zhao}, {Zhao}, {Zhu}, {Algaba},
  {Allardi}, {Amestica}, {Anczarski}, {Bach}, {Baganoff}, {Beaudoin}, {Benson},
  {Berthold}, {Blanchard}, {Blundell}, {Bustamente}, {Cappallo},
  {Castillo-Dom{\'\i}nguez}, {Chang}, {Chang}, {Chang}, {Chen}, {Chilson},
  {Chuter}, {Rosado}, {Coulson}, {Crowley}, {Derome}, {Dexter}, {Dornbusch},
  {Dudevoir}, {Dzib}, {Eckart}, {Eckert}, {Erickson}, {Everett}, {Faber},
  {Farah}, {Fath}, {Folkers}, {Forbes}, {Freund}, {Gale}, {Gao}, {Geertsema},
  {Graham}, {Greer}, {Grosslein}, {Gueth}, {Haggard}, {Halverson}, {Han},
  {Han}, {Hao}, {Hasegawa}, {Henning}, {Hern{\'a}ndez-G{\'o}mez},
  {Herrero-Illana}, {Heyminck}, {Hirota}, {Hoge}, {Huang}, {Violette
  Impellizzeri}, {Jiang}, {John}, {Kamble}, {Keisler}, {Kimura}, {Kono},
  {Kubo}, {Kuroda}, {Lacasse}, {Laing}, {Leitch}, {Li}, {Lin}, {Liu}, {Liu},
  {Lu}, {Marson}, {Martin-Cocher}, {Massingill}, {Matulonis}, {McColl},
  {McWhirter}, {Messias}, {Meyer-Zhao}, {Michalik}, {Monta{\~n}a},
  {Montgomerie}, {Mora-Klein}, {Muders}, {Nadolski}, {Navarro}, {Neilsen},
  {Nguyen}, {Nishioka}, {Norton}, {Nowak}, {Nystrom}, {Ogawa}, {Oshiro},
  {Oyama}, {Parsons}, {Pe{\~n}alver}, {Phillips}, {Poirier}, {Pradel},
  {Primiani}, {Raffin}, {Rahlin}, {Reiland}, {Risacher}, {Ruiz},
  {S{\'a}ez-Mada{\'\i}n}, {Sassella}, {Schellart}, {Shaw}, {Silva}, {Shiokawa},
  {Smith}, {Snow}, {Souccar}, {Sousa}, {Sridharan}, {Srinivasan}, {Stahm},
  {Stark}, {Story}, {Timmer}, {Vertatschitsch}, {Walther}, {Wei}, {Whitehorn},
  {Whitney}, {Woody}, {Wouterloot}, {Wright}, {Yamaguchi}, {Yu}, {Zeballos},
  {Zhang}, {Ziurys}, \& {Event Horizon Telescope Collaboration}}]{r:3c279_eht}
{Kim}, J.-Y., {Krichbaum}, T.~P., {Broderick}, A.~E., {et~al.} 2020, \aap, 640,
  A69, \dodoi{10.1051/0004-6361/202037493}

\bibitem[{{Koryukova} {et~al.}(2022){Koryukova}, {Pushkarev}, {Plavin}, \&
  {Kovalev}}]{r:scat2}
{Koryukova}, T.~A., {Pushkarev}, A.~B., {Plavin}, A.~V., \& {Kovalev}, Y.~Y.
  2022, \mnras, 515, 1736, \dodoi{10.1093/mnras/stac1898}

\bibitem[{{Kovalev} {et~al.}(2008){Kovalev}, {Lobanov}, {Pushkarev}, \&
  {Zensus}}]{r:kov08}
{Kovalev}, Y.~Y., {Lobanov}, A.~P., {Pushkarev}, A.~B., \& {Zensus}, J.~A.
  2008, \aap, 483, 759, \dodoi{10.1051/0004-6361:20078679}

\bibitem[{{Kovalev} {et~al.}(2007){Kovalev}, {Petrov}, {Fomalont}, \&
  {Gordon}}]{r:vcs5}
{Kovalev}, Y.~Y., {Petrov}, L., {Fomalont}, E.~B., \& {Gordon}, D. 2007, \aj,
  133, 1236, \dodoi{10.1086/511157}

\bibitem[{{Krasna} {et~al.}(2023){Krasna}, {Gordon}, {de Witt}, \&
  {Jacobs}}]{r:kba}
{Krasna}, H., {Gordon}, D., {de Witt}, A., \& {Jacobs}, C.~S. 2023, arXiv
  e-prints, arXiv:2306.09747, \dodoi{10.48550/arXiv.2306.09747}

\bibitem[{{Lanyi} {et~al.}(2010){Lanyi}, {Boboltz}, {Charlot}, {Fey},
  {Fomalont}, {Geldzahler}, {Gordon}, {Jacobs}, {Ma}, {Naudet}, {Romney},
  {Sovers}, \& {Zhang}}]{r:kq1}
{Lanyi}, G.~E., {Boboltz}, D.~A., {Charlot}, P., {et~al.} 2010, \aj, 139, 1695,
  \dodoi{10.1088/0004-6256/139/5/1695}

\bibitem[{{Lister} {et~al.}(2016){Lister}, {Aller}, {Aller}, {Homan},
  {Kellermann}, {Kovalev}, {Pushkarev}, {Richards}, {Ros}, \&
  {Savolainen}}]{r:mojave2}
{Lister}, M.~L., {Aller}, M.~F., {Aller}, H.~D., {et~al.} 2016, \aj, 152, 12,
  \dodoi{10.3847/0004-6256/152/1/12}

\bibitem[{{Lobanov}(1998)}]{r:lob98}
{Lobanov}, A.~P. 1998, \aap, 330, 79, \dodoi{10.48550/arXiv.astro-ph/9712132}

\bibitem[{{Ma} {et~al.}(1998){Ma}, {Arias}, {Eubanks}, {Fey}, {Gontier},
  {Jacobs}, {Sovers}, {Archinal}, \& {Charlot}}]{r:icrf1}
{Ma}, C., {Arias}, E.~F., {Eubanks}, T.~M., {et~al.} 1998, \aj, 116, 516,
  \dodoi{10.1086/300408}

\bibitem[{{MacMillan} \& {Ma}(1994)}]{r:mcm94}
{MacMillan}, D.~S., \& {Ma}, C. 1994, \jgr, 99, 637, \dodoi{10.1029/93JB02162}

\bibitem[{{Matveenko} {et~al.}(1965){Matveenko}, {Kardashev}, \&
  {Sholomitskii}}]{r:mat65}
{Matveenko}, L.~I., {Kardashev}, N.-S., \& {Sholomitskii}, G.-B. 1965, Soviet
  Radiophys., 461, 461

\bibitem[{{Mo{\'o}r} {et~al.}(2011){Mo{\'o}r}, {Frey}, {Lambert}, {Titov}, \&
  {Bakos}}]{r:moor11}
{Mo{\'o}r}, A., {Frey}, S., {Lambert}, S.~B., {Titov}, O.~A., \& {Bakos}, J.
  2011, \aj, 141, 178, \dodoi{10.1088/0004-6256/141/6/178}

\bibitem[{{Niell} {et~al.}(2018){Niell}, {Barrett}, {Burns}, {Cappallo},
  {Corey}, {Derome}, {Eckert}, {Elosegui}, {McWhirter}, {Poirier},
  {Rajagopalan}, {Rogers}, {Ruszczyk}, {SooHoo}, {Titus}, {Whitney}, {Behrend},
  {Bolotin}, {Gipson}, {Gordon}, {Himwich}, \& {Petrachenko}}]{r:vgos}
{Niell}, A., {Barrett}, J., {Burns}, A., {et~al.} 2018, Radio Science, 53,
  1269, \dodoi{10.1029/2018RS006617}

\bibitem[{{Petit} \& {Luzum}(2010)}]{r:iers2010}
{Petit}, G., \& {Luzum}, B. 2010, IERS Technical Note, 36, 1

\bibitem[{{Petrov}(2013)}]{r:obrs2}
{Petrov}, L. 2013, \aj, 146, 5, \dodoi{10.1088/0004-6256/146/1/5}

\bibitem[{{Petrov}(2015)}]{r:padel}
---. 2015, ArXiv e-prints, 1502.06678.
\newblock \url{https://arxiv.org/pdf/1502.06678}

\bibitem[{{Petrov}(2021)}]{r:wfcs}
---. 2021, \aj, 161, 14, \dodoi{10.3847/1538-3881/abc4e1}

\bibitem[{{Petrov}(2023)}]{r:sba}
---. 2023, \aj, 165, 183, \dodoi{10.3847/1538-3881/acc174}

\bibitem[{{Petrov} {et~al.}(2019){Petrov}, {de Witt}, {Sadler}, {Phillips}, \&
  {Horiuchi}}]{r:lcs2}
{Petrov}, L., {de Witt}, A., {Sadler}, E.~M., {Phillips}, C., \& {Horiuchi}, S.
  2019, \mnras, 485, 88, \dodoi{10.1093/mnras/stz242}

\bibitem[{{Petrov} {et~al.}(2009){Petrov}, {Gordon}, {Gipson}, {MacMillan},
  {Ma}, {Fomalont}, {Walker}, \& {Carabajal}}]{r:rdv}
{Petrov}, L., {Gordon}, D., {Gipson}, J., {et~al.} 2009, Journal of Geodesy,
  83, 859, \dodoi{10.1007/s00190-009-0304-7}

\bibitem[{{Petrov} \& {Kovalev}(2017)}]{r:gaia3}
{Petrov}, L., \& {Kovalev}, Y.~Y. 2017, \mnras, 471, 3775,
  \dodoi{10.1093/mnras/stx1747}

\bibitem[{{Petrov} {et~al.}(2011){Petrov}, {Kovalev}, {Fomalont}, \&
  {Gordon}}]{r:vgaps}
{Petrov}, L., {Kovalev}, Y.~Y., {Fomalont}, E.~B., \& {Gordon}, D. 2011, \aj,
  142, 35, \dodoi{10.1088/0004-6256/142/2/35}

\bibitem[{{Petrov} \& {Ma}(2003)}]{r:harpos}
{Petrov}, L., \& {Ma}, C. 2003, Journal of Geophysical Research (Solid Earth),
  108, 2190, \dodoi{10.1029/2002JB001801}

\bibitem[{{Petrov} \& {Taylor}(2011)}]{r:astro_vips}
{Petrov}, L., \& {Taylor}, G.~B. 2011, \aj, 142, 89,
  \dodoi{10.1088/0004-6256/142/3/89}

\bibitem[{{Plank} {et~al.}(2016){Plank}, {Shabala}, {McCallum},
  {Kr{\'a}sn{\'a}}, {Petrachenko}, {Rastorgueva-Foi}, \& {Lovell}}]{r:pla16}
{Plank}, L., {Shabala}, S.~S., {McCallum}, J.~N., {et~al.} 2016, \mnras, 455,
  343, \dodoi{10.1093/mnras/stv2080}

\bibitem[{{Plavin} {et~al.}(2022){Plavin}, {Kovalev}, \&
  {Pushkarev}}]{r:pla_jet}
{Plavin}, A.~V., {Kovalev}, Y.~Y., \& {Pushkarev}, A.~B. 2022, \apjs, 260, 4,
  \dodoi{10.3847/1538-4365/ac6352}

\bibitem[{{Plavin} {et~al.}(2019){Plavin}, {Kovalev}, {Pushkarev}, \&
  {Lobanov}}]{r:pla19}
{Plavin}, A.~V., {Kovalev}, Y.~Y., {Pushkarev}, A.~B., \& {Lobanov}, A.~P.
  2019, \mnras, 485, 1822, \dodoi{10.1093/mnras/stz504}

\bibitem[{{Porcas}(2009)}]{r:por09}
{Porcas}, R.~W. 2009, \aap, 505, L1, \dodoi{10.1051/0004-6361/200912846}

\bibitem[{{Pushkarev} \& {Kovalev}(2015)}]{r:scat1}
{Pushkarev}, A.~B., \& {Kovalev}, Y.~Y. 2015, \mnras, 452, 4274,
  \dodoi{10.1093/mnras/stv1539}

\bibitem[{{Rienecker} {et~al.}(2018){Rienecker}, {Suarez}, {Todling},
  {Bacmeister}, {Takacs}, {Liu}, {Sienkiewicz}, {Koster}, {Gelaro}, I., \&
  E.}]{r:geos18}
{Rienecker}, M., {Suarez}, M., {Todling}, R., {et~al.} 2018, NASA Technical
  Memorandum, 104606, 1.
\newblock \url{http://gmao.gsfc.nasa.gov/pubs/docs/tm28.pdf}

\bibitem[{{Schaer}(1999)}]{r:schaer99}
{Schaer}, S. 1999, Geod.-Geophys.~Arb.~Schweiz, Vol.~59,, 59

\bibitem[{{Sokolovsky} {et~al.}(2011){Sokolovsky}, {Kovalev}, {Pushkarev}, \&
  {Lobanov}}]{r:sok11}
{Sokolovsky}, K.~V., {Kovalev}, Y.~Y., {Pushkarev}, A.~B., \& {Lobanov}, A.~P.
  2011, \aap, 532, A38, \dodoi{10.1051/0004-6361/201016072}

\bibitem[{{Thomas} {et~al.}(2024){Thomas}, {MacMillan}, \& {Le Bail}}]{r:r1r4}
{Thomas}, C.~C., {MacMillan}, D.~S., \& {Le Bail}, K. 2024, Advances in Space
  Research, 73, 317, \dodoi{10.1016/j.asr.2023.07.020}

\bibitem[{Thomas(1980)}]{r:tho80}
Thomas, J. 1980, {JPL Publ.}, \mbox{810--005}

\bibitem[{{Xu} {et~al.}(2022){Xu}, {Savolainen}, {Anderson}, {Kareinen},
  {Zubko}, {Lunz}, \& {Schuh}}]{r:min22}
{Xu}, M.~H., {Savolainen}, T., {Anderson}, J.~M., {et~al.} 2022, \aap, 663,
  A83, \dodoi{10.1051/0004-6361/202140840}

\end{thebibliography}

\end{document}